\pdfoutput=1
\documentclass[12pt]{article}

\usepackage{amsmath,graphicx}
\usepackage{simplewick}
\usepackage{epsf}
\usepackage{graphicx,epsfig}
\usepackage{amsfonts}
\usepackage{amssymb}
\usepackage[nosort]{cite}
\usepackage{setspace}

\def\bk{{\bf k}}
\def\bp{{\bf p}}

\def\CL{{\cal L}}
\def\CO{{\cal O}}

\def\high{\vphantom{\Biggl(}\displaystyle}

\def\mpl{M_{\rm P}}

\def\mhor{m_{\rm h}}

\makeatletter
\renewcommand\section{\@startsection {section}{1}{\z@}%
                                 {-3.5ex \@plus -1ex \@minus -.2ex}%
                                   {2.3ex \@plus.2ex}%
                                   {\normalfont\large\bfseries}}
\renewcommand\subsection{\@startsection{subsection}{2}{\z@}%
                                   {-3.25ex\@plus -1ex \@minus -.2ex}%
                                     {1.5ex \@plus .2ex}%
                                     {\normalfont\bfseries}}
\renewcommand\subsubsection{\@startsection{subsubsection}{3}{\z@}%
                                   {-3.25ex\@plus -1ex \@minus -.2ex}%
                                     {1.5ex \@plus .2ex}%
                                     {\normalfont\itshape}}
\makeatother

\newcommand{\Letter}{
\setlength{\textwidth}{16.5cm}
   \setlength{\textheight}{23cm}
    \hoffset=-0.5in
\voffset=-2.1cm }

\Letter

\begin{document}
\newcommand{\be}{\begin{equation}}
\newcommand{\ee}{\end{equation}}
\newcommand{\bea}{\begin{eqnarray}}
\newcommand{\eea}{\end{eqnarray}}
\newcommand{\barr}{\begin{array}}
\newcommand{\earr}{\end{array}}
\def\bal#1\eal{\begin{align}#1\end{align}}

\newcommand{\nc}{\newcommand}
\nc{\ba}{\begin{eqnarray}}
\nc{\ea}{\end{eqnarray}}
\newcommand{\calR}{{\cal{R}}}
\newcommand{\calP}{{\cal{P}}}
\newcommand{\calH}{{\cal{H}}}
\newcommand{\calI}{{\cal{I}}}

\thispagestyle{empty}
\begin{flushright}
\end{flushright}

\vspace*{0.3in}
\begin{spacing}{1.1}

\begin{center}
{\large \bf Quantum Primordial Standard Clocks}

\vspace*{0.3in} {Xingang Chen$^{1,2}$, Mohammad Hossein Namjoo$^{1,2}$, and Yi Wang$^3$}
\\[.3in]
{\em
$^1$Department of Physics, The University of Texas at Dallas, Richardson, TX 75083, USA \\
$^2$ Institute for Theory and Computation, Harvard-Smithsonian Center for Astrophysics,\\
60 Garden Street, Cambridge, MA 02138, USA\\
$^3$Department of Physics, The Hong Kong University of Science and Technology,\\
Clear Water Bay, Kowloon, Hong Kong, P.R.China} \\[0.3in]

\end{center}

\begin{center}
{\bf
Abstract}
\end{center}
\noindent
In this paper, we point out and study a generic type of signals existing in the primordial universe models, which can be used to model-independently distinguish the inflation scenario from alternatives. These signals are generated by massive fields that function as standard clocks. The role of massive fields as standard clocks has been realized in previous works. Although the existence of such massive fields is generic, the previous realizations require sharp features to classically excite the oscillations of the massive clock fields. Here, we point out that the quantum fluctuations of massive fields can actually serve the same purpose as the standard clocks. We show that they are also able to directly record the defining property of the scenario type, namely, the scale factor of the primordial universe as a function of time $a(t)$, but through shape-dependent oscillatory features in non-Gaussianities. Since quantum fluctuating massive fields exist in any realistic primordial universe models, these quantum primordial standard clock signals are present in any inflation models, and should exist quite generally in alternative-to-inflation scenarios as well. However, the amplitude of such signals is very model-dependent.

\vfill

\newpage
\setcounter{page}{1}

\tableofcontents

\newpage

\section{Introduction}
\label{Sec:Introduction}
\setcounter{equation}{0}

In the past decades, through interactions between observations and theories, the Standard Model of cosmology has been established better than ever. With more abundant data coming from Cosmic Microwave Background (CMB) and Large Scale Structure (LSS) in the near future, there are potentially different types of signals in data that could provide information beyond the Standard Model and allow us to address many deeper questions. In this paper we are interested in one of the most important questions in the early universe cosmology, namely, how to observationally distinguish the inflation scenario \cite{Guth:1980zm} from the alternatives. Experimental data containing information on the primordial fluctuations determines how deeply we could address this question. Although many possible beyond-Standard-Model signals could provide invaluable information on model details within a scenario, unfortunately only very few of them could be used to {\em model-independently} distinguish different scenarios.\footnote{Nonetheless, we stress the current status that the inflation scenario is much better established than any other alternative scenarios. See \cite{Brandenberger:2012zb} for recent discussions. As we will see, the approach in our paper on this model-building aspect is phenomenological.}

A well-known example is the tensor mode of the primordial fluctuations \cite{Grishchuk:1974ny}, which may be observed as the primordial B-mode polarization in CMB \cite{Seljak:1996ti}. This signal records the magnitude of the Hubble rate in the primordial universe, and thus is able to at least model-independently distinguish inflation from some of the alternative scenarios with much more slowly varying scale factors.\footnote{Having that said, some alternative scenarios still give similar prediction of gravitational waves. For example, both matter bounce and string gas cosmology predict nearly scale invariant spectrum of gravitational waves, and one needs to rely on details of the gravitational wave spectrum, as well as other observables, to distinguish such scenarios \cite{Wang:2014kqa, Quintin:2015rta}.}

Another example that has been proposed recently is to make use of the oscillation of massive fields as the standard clocks \cite{Chen:2011zf,Chen:2014cwa} to directly record the time-evolution of the scale factor of the primordial universe $a(t)$ -- the defining property of a specific scenario. In previous works, this idea has been realized in terms of the classical process \cite{Chen:2011zf,Chen:2014cwa}. Namely, some kind of sharp features are necessary in the models to classically excite the oscillation of the massive clock fields. The ticks of the clock are recorded as special scale-dependent oscillatory features in the density perturbations, carrying direct information of $a(t)$. Although the existence of the required massive fields should be general in any realistic model, the sharp features may not exist in any model despite of the fact that there are some well-motivated examples \cite{Chen:2014cwa}. We call this case the ``classical primordial standard clock".

In this paper, we point out and explain that the same idea can actually be realized by quantum fluctuations of these massive fields in the time-dependent background, which is now completely general.

If the mass of a candidate clock field is large enough, there is always a regime during its evolution in which this mass is larger than both the physical wavenumber of the clock field and the mass scale of the horizon. We call this regime the classical regime. In the classical regime, the quantum-mechanically-originated oscillations of these massive fields behave very much like those in the classical standard clock case, in the sense that these fields are approximately homogeneous over a region much larger than their Compton scales and they are oscillating as if in an approximately flat spacetime. For example, in the expansion scenarios, this classical regime happens in a relatively later stage during the evolution of the clock field, when the wavelength of the clock field gets stretched so that its physical wavenumber falls below its mass, and at the same time when the horizon scale is still below the mass; in the contraction scenarios, this classical regime happens in an earlier stage of the evolution, before both the wavelength of the clock field and the size of the horizon get contracted and hence their mass scales exceed the mass of the clock field.

Once this is realized, the dynamics of recording the scale factor evolution $a(t)$ becomes similar to that in the classical standard clock case. The ticks of these classical-like oscillations of the clock field get imprint in the correlation functions through the resonance mechanism \cite{Chen:2008wn,Flauger:2009ab,Flauger:2010ja,Chen:2010bka}. However, unlike the classical standard clock case, the quantum fluctuation of the massive field happens persistently and is not a source of scale-invariance breaking. Interestingly, this crucial difference determines how the clock signals in the quantum case get subtly encoded in the density perturbations. They are encoded not as scale-dependent oscillatory features, but as {\em shape-dependent} oscillatory features in {\em non-Gaussianities}.
We call this case the ``quantum primordial standard clock".

For the inflation scenario, the effects of quantum fluctuations of massive field with mass of order the Hubble parameter $H$ or larger on density perturbations have been studied to some extent \cite{Chen:2009we,Baumann:2011nk,Assassi:2012zq,Chen:2012ge,Noumi:2012vr,Arkani-Hamed:2015bza,Sefusatti:2012ye,Norena:2012yi,Gong:2013sma,Emami:2013lma,Jackson:2012fu,Craig:2014rta,Dimastrogiovanni:2015pla,Schmidt:2015xka}. The most striking feature is the distinctive scaling behavior of the non-Gaussianities as a function of momentum ratios. In the squeezed limit of the three-point correlation functions, the shape of non-Gaussianity $S$ scales as \cite{Chen:2009we,Noumi:2012vr,Arkani-Hamed:2015bza},
\bea
S \propto  \left( \frac{k_{\rm long}}{k_{\rm short}} \right)^{\alpha}
P_s(\cos\theta) ~,
\eea
where $\theta$ is the angle between the long and short mode ($\bk_{\rm long}$ and $\bk_{\rm short}$) and $s$ is the spin of the massive field.
For this work, the most important feature in this formula is the scaling power $\alpha$, which is determined by the mass $m$ of the field,
\bea
\alpha = \frac{3}{2} \pm i \sqrt{\frac{m^2}{H^2} - \frac{9}{4}} ~.
\eea
The shape-dependent oscillation for the $m>3H/2$ case is emphasized in \cite{Arkani-Hamed:2015bza} and interpreted as a consequence of the quantum interference between the inflaton and the massive field. In this paper, we point out another physical significance of this result -- it is the quantum mechanical realization of the primordial standard clock and the signal directly encodes the scale factor evolution of the inflation scenario, $a(t) \sim \exp(Ht)$. This interpretation is important, because as mentioned, there are different kinds of beyond-Standard-Model signals from inflation models, and it is crucial to realize which ones can be used to model-independently distinguish the inflation scenario from the alternatives. We have briefly explained the reasons behind this realization, and in the main text we shall illustrate them in details. For this purpose, it is also necessary to go beyond the inflation scenario and we shall study the effect of the quantum standard clocks in the alternative scenarios. We will show explicitly how their scale factor evolution $a(t)$ is encoded in the clock signals.

Massive fields are present in any UV-completed primordial universe models; they quantum fluctuate in any time-dependent backgrounds; and they couple to any other fields at least through gravity. Therefore, the signals of quantum primordial standard clocks are present in any inflation models, and should exist quite generally in the alternative-to-inflation scenarios as well if the model-building would become more complete. However, the amplitude of such signals is highly model-dependent and its order of magnitude spans a wide range. In the gravitational coupling case, the amplitude is suppressed by the slow-roll parameters for inflation models \cite{Chen:2009we}; and if the mass of the clock field is much higher than the horizon mass, we expect additional Boltzmann suppression factors \cite{Arkani-Hamed:2015bza}. On the other hand, for clock fields with mass of order the horizon scale (e.g.~$\CO(H)$ in inflation) and for non-gravitational couplings, the signals may be potentially observable in the future \cite{Chen:2009we,Baumann:2011nk}. The generic existence of the signals and the variety of their amplitudes are very similar to the situations for the tensor mode.

This paper is organized as follows.
In Sec.~\ref{Sec:Quantum_massive}, we study quantum fluctuations of the massive scalar field in arbitrary time-dependent background and emphasize the classical regime during which the massive field can serve as a standard clock. We use Fig.~\ref{Fig:scales} to qualitatively illustrate the relations of different scales in different scenarios, and the appearance of the classical regime in each scenario.
In Sec.~\ref{Sec:QSC}, we first briefly review the mechanism of the classical standard clock, and then explain and summarize the main features and results of the quantum standard clock. We show how the scale factor evolution $a(t)$ is directly encoded in the squeezed-limit bispectrum through an example.
In Sec.~\ref{Sec:Inflation} and \ref{Sec:Alternatives}, we show the details of the quantum standard clocks in the inflation scenario and the alternative scenarios, respectively.
In Sec.~\ref{Sec:Amplitude}, we demonstrate through examples the highly model-dependent range of the amplitude of the clock signals in the inflation scenario.
Sec.~\ref{Sec:Discussions} is conclusions and discussions.

\section{Quantum fluctuations of massive fields}
\label{Sec:Quantum_massive}
\setcounter{equation}{0}

For generality and simplicity, as in \cite{Chen:2011zf} we take the phenomenological approach and describe different kinds of time-dependent backgrounds of the primordial universe using a simple power-law function,
\bea
a(t) = a_0 \left( \frac{t}{t_0} \right)^p ~,
\label{Power-law bkgd}
\eea
where $a_0$ and $t_0$ are constants.
This time-dependent background has a horizon size $|t/(1-p)|$.
The requirement that the quantum fluctuations exit the horizon fixes the direction of $t$ given the value of $p$. There are at least four interesting possibilities: $|p|>1$ corresponds to the fast-expansion scenario, namely inflation; $0<p\sim \CO(1)<1$ corresponds to the fast-contraction scenario; $0<p\ll 1$ corresponds to the slow-contraction scenario; and $-1 \ll p <0$ corresponds to the slow-expansion scenario.\footnote{We comment that, while the value of $p$ is sufficient to define the scenario type, the terms ``fast" and ``slow" for different $p$-value ranges are named empirically. For example, in the available examples, the $0<p\ll 1$ scenario typically has very small Hubble parameter and the $0<p\sim \CO(1)<1$ scenario has large Hubble parameter, although this does not have to be the case just in terms of the phenomenological parameterization \eqref{Power-law bkgd}.}
The time $t$ runs from $0$ to $+\infty$ for $p>1$ and from $-\infty$ to $0$ for all other $p$ values. The conformal time $\tau$, which in this background is related to $t$ by
\bea
a\tau=\frac{t}{1-p} ~,
\label{t_tau_relation}
\eea
always runs from $-\infty$ to $0$.

We now study the behavior of massive fields in this general background. We study scalar fields in this paper. Generalization to higher spin massive fields should also be interesting.

The equation of motion of a massive field quantum fluctuation, $\delta\sigma$, in arbitrary time-dependent background is
\bea
\ddot{\delta\sigma} + 3 H \dot{\delta\sigma} - \frac{1}{a^2} \partial_i^2 \delta\sigma + m^2 \delta\sigma =0 ~,
\label{Sigma_EoM_in_t}
\eea
where dots denote derivatives with respect to time $t$; or in terms of the conformal time,
\bea
\delta\sigma'' + 2H a \delta\sigma' - \partial_i^2 \delta\sigma + m^2 a^2 \delta\sigma =0 ~,
\label{Sigma_EoM_in_tau}
\eea
where primes denote derivatives with respect to the conformal time $\tau$. We will be interested in the cases where the mass is larger than the energy scale of the horizon,
\bea
m > \mhor ~,
\label{Mass_cond}
\eea
where
\bea
\mhor \equiv \left| \frac{1-p}{t} \right| = \frac{1}{a|\tau|} ~.
\label{mhor_def}
\eea

Even with the simple power-law (\ref{Power-law bkgd}), this differential equation cannot be solved analytically for arbitrary $p$. But the analytical expressions in the following limits are useful.

\subsection{Relativistic regime}

Denote the Fourier mode of $\delta\sigma$ as $v_\bk$. Since the background is isotropic, $v_\bk$ only depends on the magnitude of the mode $\bk$. When the physical wavelength of a mode is short enough to satisfy
\bea
k/a \gg m ~,
\label{UV_condition1}
\eea
the EoM (\ref{Sigma_EoM_in_tau}) is approximately
\bea
v_k''+ 2 H a v_k' + k^2 v_k =0 ~.
\eea
In the limit
\bea
|k\tau| \gg {\rm Min}(p,1) ~,
\label{UV_condition2}
\eea
we can ignore higher order terms and the solution of this equation is approximately
\bea
v_k \to \frac{-i}{\sqrt{2k}} a_0^{-1} \left( \frac{\tau}{\tau_0} \right)^{\frac{p}{p-1}} e^{-ik\tau} ~,
\label{UV_limit}
\eea
where the normalization is fixed by the canonical quantization condition. We have also chosen the BD vacuum and we will further discuss this choice at the end of this section.

Using (\ref{t_tau_relation}) we can see that the condition (\ref{UV_condition2}) implies that the modes are subhorizon $k/a \gg |(1-p)/t|$. Nonetheless, for the massive field that satisfies (\ref{Mass_cond}), the condition (\ref{UV_condition1}) is stronger than (\ref{UV_condition2}).

\subsection{Classical regime}
\label{Sec:classical_regime}

When the physical wavelength of a mode is large enough to satisfy
\bea
k/a \ll m ~,
\label{classical_condition}
\eea
the EoM becomes approximately
\bea
\ddot v_k + 3 H \dot v_k + m^2 v_k =0 ~.
\eea
This equation can also be solved assuming (\ref{Power-law bkgd}), and we will be interested in the large $|t|$ behavior,
\bea
v_k \to \left( \frac{t}{t_{k}} \right)^{-3p/2}
\left( c_+ e^{-imt} + c_- e^{imt} \right) ~,
\label{classical_limit}
\eea
where the higher order terms can be neglected if
\bea
\left| mt \right| \gg {\rm Max}(p^2,1) ~.
\label{Mass_cond2}
\eea
The parameter $t_k$ is the time at which $k/a|_{t=t_k} = m$, and the quantization condition requires
\bea
2m a^3(t_k) \left( |c_+|^2 - |c_-|^2 \right) = 1 ~.
\label{alpha_quan_cond}
\eea

\begin{figure}[htbp]
  \centering
  \includegraphics[width=0.65\textwidth]{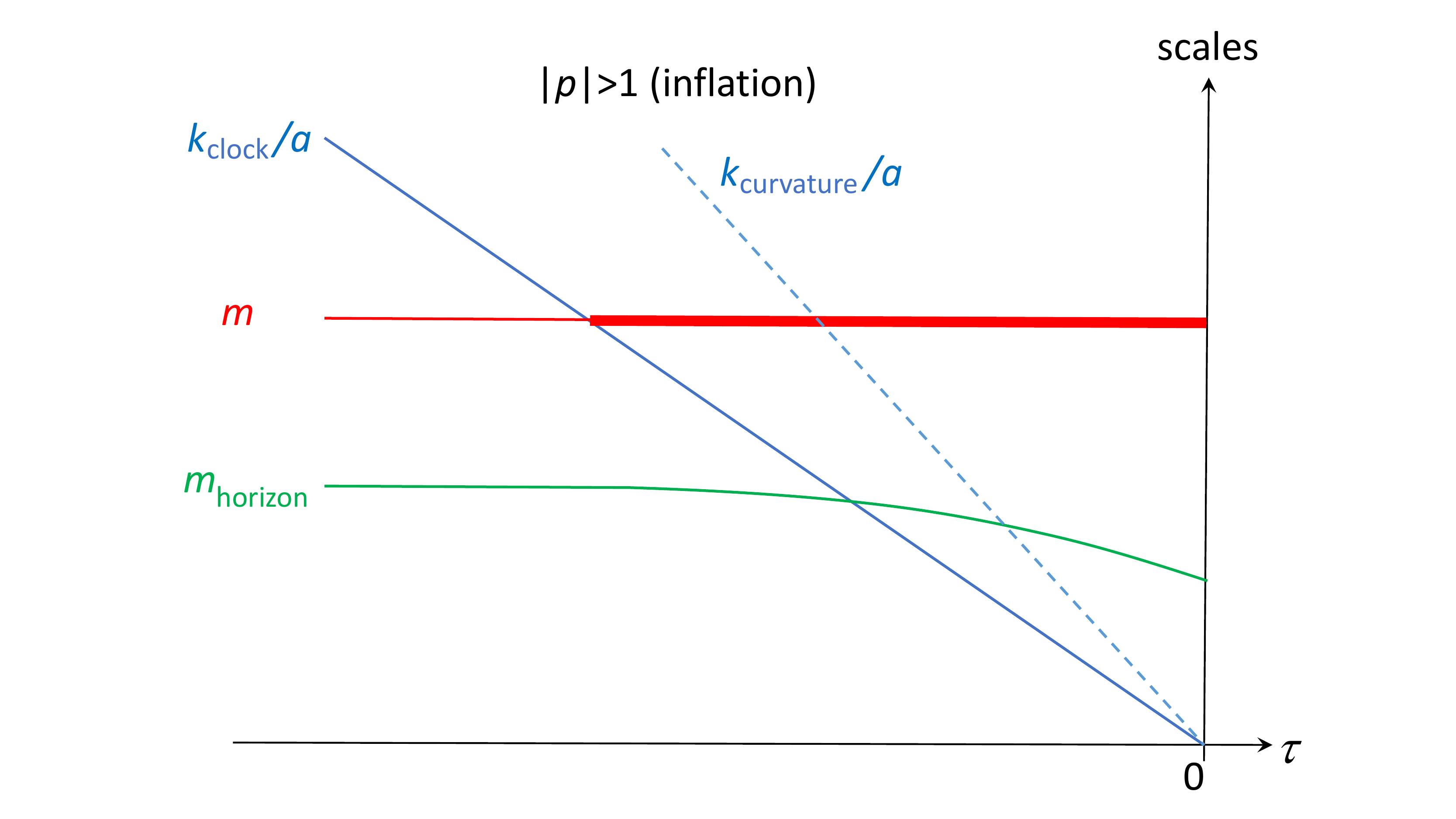}
  \includegraphics[width=0.65\textwidth]{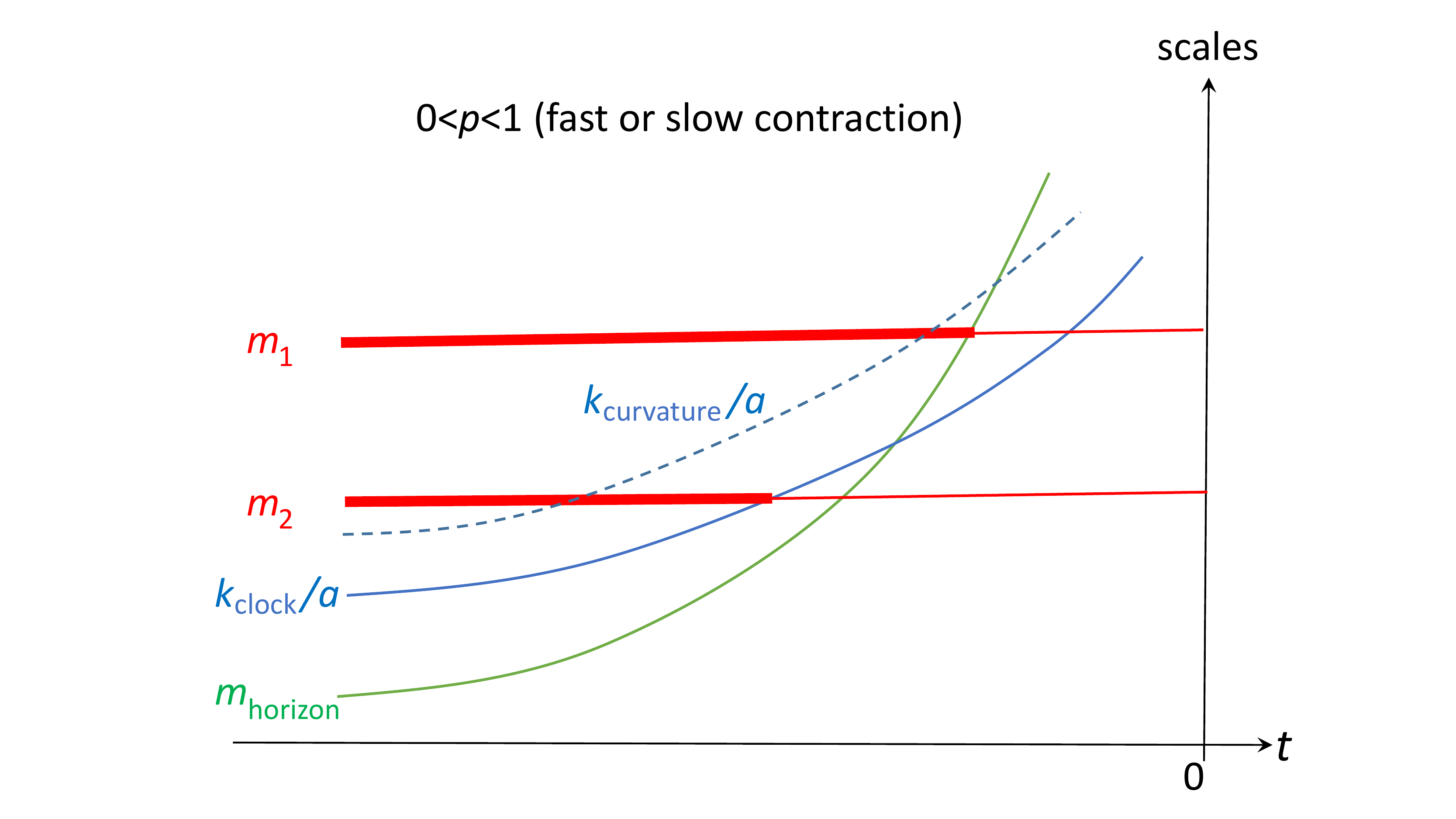}
  \includegraphics[width=0.65\textwidth]{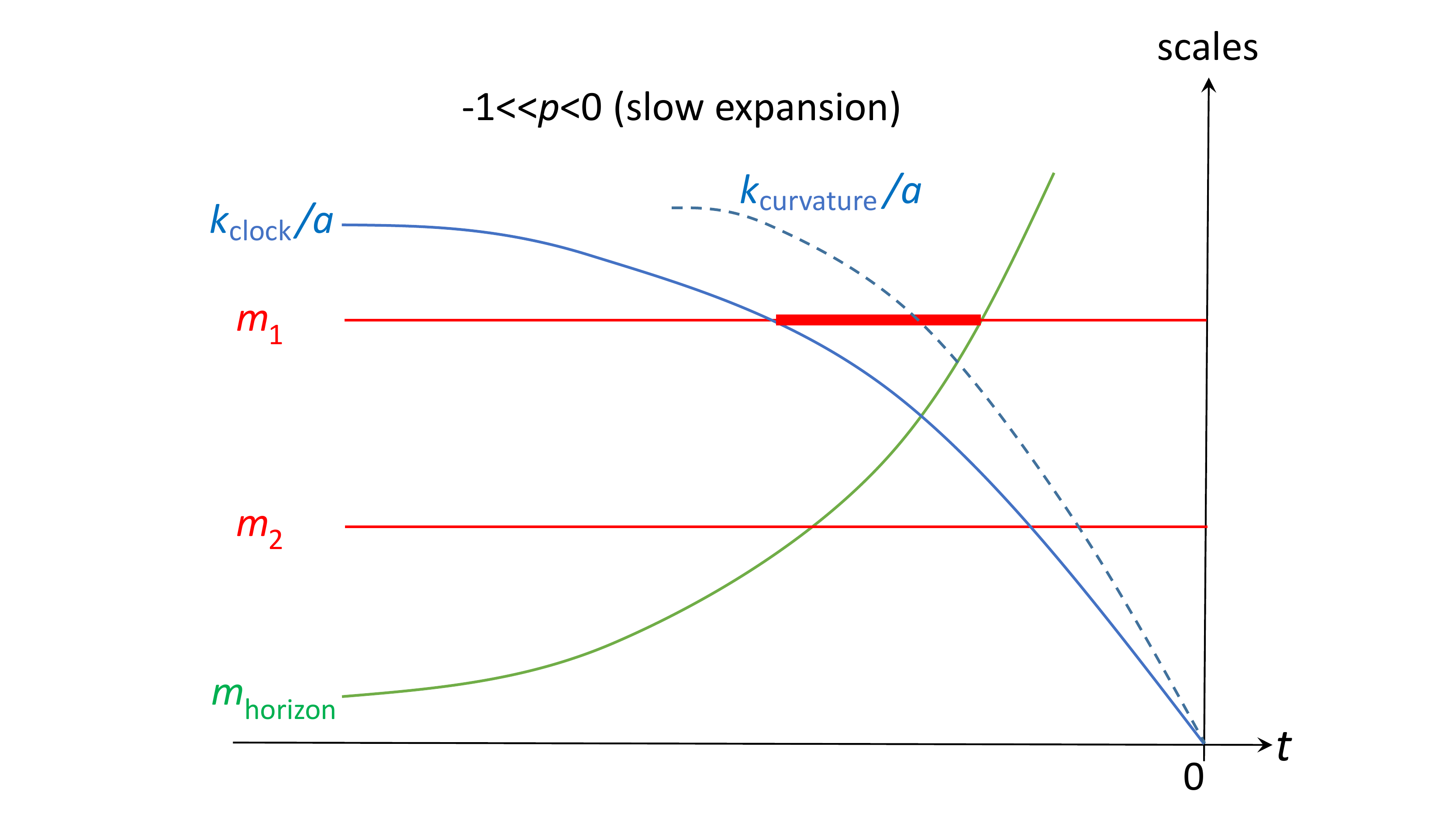}
\caption{\small These figures illustrate the conditions for a massive field to be qualified as a quantum standard clock field in different scenarios. We qualitatively sketch the evolution of four scales (the constant mass of the massive field $m$, the physical wavenumber of the massive field $k_{\rm clock}/a$, the mass of the horizon scale $m_{\rm horizon}$, and the physical wavenumber of the massless curvature scalar mode $k_{\rm curvature}/a$) as functions of the conformal time $\tau$ or real time $t$ in the inflation, fast (or slow) contraction, and slow expansion scenarios, respectively.
Two different cases of $m$ values are shown in some scenarios. The {\bf thick lines} indicate the classical regime during which the massive field can be used as a standard clock. When the dashed lines intersect with the thick lines, the resonance between the curvature scalar mode and the clock field happens, and in all cases we find $k_{\rm clock} < k_{\rm curvature}$. At the resonance, $k_{\rm curvature}/a$ is always at the sub-horizon scales, $k_{\rm clock}/a$ can be at either the sub- or super-horizon scales.}
\label{Fig:scales}
\end{figure}

To summarize, we define the classical regime to be the regime where the mass of the massive field is larger than both the physical wavenumber of the field and the horizon mass scale.
The behavior (\ref{classical_limit}) holds in the classical regime where both (\ref{classical_condition}) and (\ref{Mass_cond2}) are satisfied. The former condition makes the field oscillate homogeneously over a region much larger than the Compton scale $\sim m^{-1}$. As we will see, $m^{-1}$ is the typical time scale in the generation mechanism of the clock signals. The latter makes the massive field oscillate with frequency $m$ as if in the approximately flat spacetime.
Note that the condition (\ref{Mass_cond}) is more general than \eqref{Mass_cond2} and is in principle enough for our purpose; in particular for the inflation case with $p\gg 1$, \eqref{Mass_cond2} is too restrictive and \eqref{Mass_cond} is sufficient.
Nonetheless, the condition (\ref{Mass_cond2}) will be used for the alternative-to-inflation scenarios because it leads to a simpler approximation (\ref{classical_limit}).

This classical-like behavior reminds us of that of the classical standard clock, and is essential for the quantum fields working as standard clocks.
Note that this is referred to as classical-like because the massive fields are not really created classically.

In Fig.~\ref{Fig:scales}, we sketch the evolution of several important scales in four scenarios.
In the caption, we also emphasize several important properties that can be observed from this figure.

As we can see from Fig.~\ref{Fig:scales}, for expansion scenarios, since the clock field originates from the relativistic ($k/a>m>\mhor$) regime, the BD vacuum is the natural initial condition to start these quantum fluctuations. The coefficients $c_{+,-}$ are then determined by the BD vacuum.
Nevertheless, as we will see later, although for the expansion scenarios we regard this BD state as the natural initial condition for the clock field, the most important property of the clock signal does not depend on this choice.

For contraction scenarios, the clock field originates from the classical ($k/a<m$ and $m>\mhor$) regime.
Only later on it evolves into the $k/a>m$ regime which may or may not satisfy the condition $m>\mhor$ (see Fig.~\ref{Fig:scales}).
So the choice of \eqref{UV_limit} may not be natural at all, and the coefficients $c_{+,-}$ are more arbitrary as long as the condition \eqref{alpha_quan_cond} is satisfied.\footnote{In fact, unlike the inflation scenario, since the contraction scenarios may or may not have an efficient way to get rid of the classical relics, even the classical origin of the clock field oscillation may be natural. Nonetheless, in this paper we focus on the minimal condition, namely the quantum fluctuations.}

\section{Quantum standard clocks}
\setcounter{equation}{0}
\label{Sec:QSC}

\subsection{Mechanism of the standard clock}

We start by briefly reviewing how the standard clock is able to record the time evolution of the background scale factor \cite{Chen:2011zf,Chen:2014cwa} in the classical standard clock case.
The classical oscillatory behavior of massive fields is used as the standard clock. These massive fields, which are heavier than the mass scale of the horizon and homogeneous over a region much larger than $1/m$, oscillate in standard ways in arbitrary time-dependent background. The most important property is its approximately constant frequency, $\sim e^{\pm imt}$, which can be regarded as a clock generating ticks.
The ticks of the standard clock get imprinted in the density perturbations through the resonance mechanism \cite{Chen:2008wn,Flauger:2009ab,Flauger:2010ja,Chen:2010bka}.\footnote{In some special cases, the ticks may also be imprinted in a non-resonant way. We will discuss this in Sec.~\ref{Sec:non-T-ordered} and Appendix \ref{App:resonance_inflation}.}
This mechanism works in the following way.
At subhorizon, the quantum fluctuations of the curvature scalar mode approach to those in the Minkowski limit. For example, if we choose the BD vacuum the mode oscillates as $\sim e^{-iK \tau}$, in which $K$ is the wavenumber.
The physical frequency of the mode changes with time due to the expansion or contraction of the background.
When this frequency coincides with that of the clock, resonance between the mode and the background happens, generating schematically the following contribution to the correlation function,
\bea
\int_{\tau_{\rm begin}}^{\tau_{\rm end}} d\tau g(t) e^{imt} e^{-iK\tau}
\to g(t_*) \int_{\tau_{\rm begin}}^{\tau_{\rm end}} d\tau e^{imt - iK\tau}
~,
\eea
where $g(t)$ represents some functions whose time dependence is much weaker than that in the other two factors, and $[\tau_{\rm begin}, \tau_{\rm end}]$ denotes the period in which the resonance happens. The resonance happens if
\bea
\left. \frac{d}{d\tau} \left( mt-K\tau \right) \right|_{t=t_*}=0 ~,
\eea
namely,
\bea
a(t_*) = K/m ~.
\eea
The integration is proportional to
\bea
\exp\left(imt_* - iK\tau_*\right) =
\exp \left( -i\frac{p}{1-p} m ~t(K/m) \right) ~,
\label{phase_behavior}
\eea
where we have used the relation (\ref{t_tau_relation}) for the power-law background. Notice that the function $t(a)$ in (\ref{phase_behavior}) is the inverse function of $a(t)$, so the scale factor evolution is directly recorded in the phase of this oscillatory signal and manifests as the relative spacing between the ticks of the signal in the $K$-space. We call this signal the {\em clock signal}.

The more precise formula that we often use is the following,
\bal
&\int_{\tau_{\rm begin}}^{\tau_{\rm end}} d\tau g(t) e^{imt} e^{-iK\tau}
\nonumber \\
\to & \sqrt{2\pi} g(t_*) \left( \frac{m}{|H_{k_0}|} \right)^{1/2} K^{-1}
\left( \frac{K}{k_0} \right)^{1/2p}
\exp\left[ -i\frac{p^2}{1-p} \frac{m}{H_{k_0}} \left( \frac{K}{k_0} \right)^{1/p} \mp i\frac{\pi}{4} \right] ~,
\label{resonance_formula}
\eal
where $k_0$ is the mode that satisfies $k_0/a=m$ at some time $t_0$, and $H_{k_0}=p/t_0$ is the Hubble parameter at $t_0$. The $\mp$ applies when $H_0$ is negative/positive.

Away from the resonance point, mostly the integrand is oscillating rapidly with time and does not give significant contribution to the integration. At or outside the boundaries of this integral, especially at the $\tau_{\rm end}$ end, in some cases there may be a finite contribution; or in other cases, if the function $g(t)$ is singular enough, the integrand can be divergent and needs to be cutoff or regulated by means of model-building which may leave some finite contribution to the integral too. In such cases, these contributions do not change the clock signal but add some overall shift.

Previous examples of standard clocks make use of the classical clocks \cite{Chen:2011zf,Chen:2014cwa}. They rely on some sharp process that excites the classical oscillation of a massive field. This process generates two different kinds of oscillatory features in the density perturbations, namely the sinusoidal sharp feature signal and resonance clock signal. These two kinds of features are smoothly connected to each other with specific patterns, and can be used as an identity for the standard clock signal in contrast to, for example, the resonance signals generated by the periodic features in the potential \cite{Chen:2008wn,Flauger:2009ab,Flauger:2010ja,Chen:2010bka}. (See \cite{Chen:2014cwa} for a summary of other typical characteristics of the classical standard clocks.)
The sinusoidal sharp feature signal is the one associated with the sharp feature.
The resonance clock signal is the one induced by the subsequent oscillation of the massive field, and it records the time evolution of $a(t)$ as illustrated above.

There are at least two universal properties that make the clock signals ``standard'':
\begin{itemize}
\item
Massive fields with mass larger than the horizon scale oscillate in standard ways in any time-dependent background. These include not only their most important oscillatory behavior, but also the typical time-dependence of the amplitudes which can also be used as auxiliary evidences.

\item
The quantum fluctuations of the curvature scalar field, which are the source of the density perturbations, are at the subhorizon scales when the clock signal is generated. The behavior of such fluctuations approaches to that in the Minkowski limit and therefore is universal for different primordial universe scenarios. This is very different from their highly model-dependent superhorizon behaviors.

\end{itemize}

\subsection{Quantum standard clocks}
\label{Sec:QuantumSC}

In this paper we propose to use quantum fluctuations of the massive field as the standard clock. In the time-dependent background, massive fields always fluctuate quantum-mechanically. We have noticed in Sec.~\ref{Sec:classical_regime} that, in the classical regime, the behavior of these quantum fluctuations resembles closely that of the classical oscillation. Besides sharing the above two universal properties, quantum standard clock exists in the absence of any sharp features. It only relies on the assumption of the existence of some heavy fields, which should be true in any realistic models.

To illustrate the mechanism and properties of the quantum standard clocks in arbitrary background, we use some simple examples of coupling terms between the curvature scalar perturbation $\zeta$ (see \cite{Maldacena:2002vr} for definition) and the clock field perturbation $\delta\sigma$. For example, we use
\bea
\CL_3 \sim c_3 a^3 \dot\zeta^2 \delta\sigma
\eea
for the cubic coupling and
\bea
\CL_2 \sim c_2 a^3 \dot\zeta \delta\sigma
\eea
for the bilinear coupling, where $c_{2,3}$ are some couplings whose time-dependence is weaker than $1/m$. The existence of these couplings require some non-minimal (i.e.~non-gravitational) couplings between the inflaton and the massive field in the model; and such an inflation model example can be found in \cite{Chen:2009we}. We discuss the gravitational coupling case in Sec.~\ref{Sec:Discussions}.

In most of this work, we do not pay much attention to the magnitude of the non-Gaussianity which we will later demonstrate to be very model-dependent. We concentrate on the momentum-shape dependence of the non-Gaussianity.

\begin{figure}[h]
  \centering
  \includegraphics[width=0.25\textwidth]{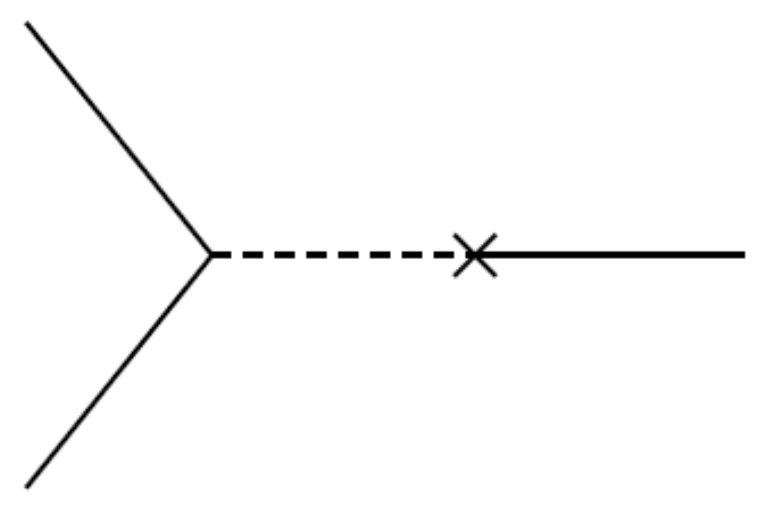}
\caption{\small An example of three-point function. The solid lines represent the curvature scalar mode, the dashed line represents the massive clock field.}
\label{Fig:3pt_vertex}
\end{figure}

Using the in-in formalism (see \cite{Chen:2010xka,Wang:2013eqj} for reviews), these two coupling terms give the following contribution to the three-point correlation function of the curvature perturbation (see Fig.~\ref{Fig:3pt_vertex}),
\bal
\langle \zeta^3 \rangle'
&\supset
\int_{t_0}^t d\tilde t_1 \int_{t_0}^t dt_1
\langle 0| H_I(\tilde t_1) \zeta_I^3 H_I(t_1) |0\rangle'
-2 {\rm Re} \left[ \int_{t_0}^t dt_1 \int_{t_0}^{t_1} dt_2
\langle 0| \zeta_I^3 H_I(t_1) H_I(t_2) |0\rangle' \right]
\nonumber \\
&= 2 u_{k_3}^* u_{k_1} u_{k_2}|_{\tau=0}
\left( \int_{-\infty}^0 d\tilde\tau c_2 a^3 v_{k_3} u'_{k_3} \right)
\left( \int_{-\infty}^0 d\tau c_3 a^2 v_{k_3}^* u'^{*}_{k_1} u'^{*}_{k_2} \right)
\label{Term1} \\
& -2u_{k_3} u_{k_1} u_{k_2}|_{\tau=0}
\nonumber \\
&\times \left[ \int_{-\infty}^0 d\tau_1 c_2 a^3 v_{k_3} u'^*_{k_3}
\int_{-\infty}^{\tau_1} d\tau_2 c_3 a^2 v_{k_3}^* u'^{*}_{k_1} u'^{*}_{k_2} + \int_{-\infty}^0 d\tau_1 c_3 a^2 v_{k_3} u'^{*}_{k_1} u'^{*}_{k_2}
\int_{-\infty}^{\tau_1} d\tau_2 c_2 a^3 v^*_{k_3} u'^*_{k_3} \right]
\label{Term2} \\
& + {\rm c.c.}+ {\rm 2~perm.} ~,
\nonumber
\eal
where we denote the mode functions for $\zeta$ and $\delta\sigma$ as $u_k$ and $v_k$, respectively; the prime in $\langle \dots \rangle'$ means that we ignore the common factor $(2\pi)^3 \delta (\sum_i \bp_i)$ in this expression. The ``2 perm." stands for two terms with the momentum permutation $k_1\leftrightarrow k_3$ and $k_2 \leftrightarrow k_3$, respectively.

Let us first look at the term (\ref{Term1}). The resonance between different mode functions only happens for the integral inside the second parenthesis. Analytical expression is possible if we look at
the classical regime and
the squeezed ($k_3/k_{1,2}\to 0$) limit, in which the massive clock field has the wavenumber $k_3$.
In the classical regime, the massive field oscillates like the classical mode, described by (\ref{classical_limit}).
For the squeezed configuration, the wavenumber of the massive mode is smaller than that of the two curvature scalar modes, and according to Fig.~\ref{Fig:scales},
this oscillation provides a fixed frequency which at some point coincides with the frequency of the two shorter curvature scalar modes as their frequency decreases (in the expansion scenarios) or increases (in the contraction scenarios).
This resonance happens when the curvature scalar modes $u_{k_1}$ and $u_{k_2}$ are well within the horizon and described by their behavior in the Minkowski limit,
\bea
u_k \to \frac{1}{a\sqrt{4\epsilon k}} e^{-ik\tau} ~,
\label{curvature_BD}
\eea
where $\epsilon \equiv -\dot H/H^2$.
We emphasize that the curvature scalar mode always starts in the subhorizon region quantum mechanically and then exits the horizon in all scenarios.
The resonance always happens when the curvature scalar mode is still in the sub-horizon region.
Therefore, the Minkowski behavior \eqref{curvature_BD} is universal for all scenarios and this makes general analyses possible.
On the other hand, although in this paper for definiteness we always choose the BD vacuum for the curvature scalar mode, this choice is not essential for the clock signal and can be relaxed.
Note that in the same squeezed limit the resonance does not happen for the other two terms with the momentum permutations.

Note that the above conclusions only rely on the general properties in Sec.~\ref{Sec:Quantum_massive} and Fig.~\ref{Fig:scales}, and qualitatively summarize how the quantum clock signals are generated in general.

Using these approximations, the integral in the second bracket of (\ref{Term1}) becomes
\bea
\propto
c_- k_1
\int_{\tau_{\rm begin}}^{\tau_{\rm end}} d\tau \left( \frac{t}{t_{k_3}} \right)^{-3p/2}
e^{-imt} e^{2i k_1 \tau} ~,
\label{clock_integral}
\eea
where we have ignored the coefficients that are independent of $k_3$ and $k_1$. The $\tau_{\rm begin}$ and $\tau_{\rm end}$ denote the time period in which the massive field satisfies the condition of being a quantum clock field, as illustrated in Fig.~\ref{Fig:scales}. According to the convention in Sec.~\ref{Sec:classical_regime}, $t_{k_3}$ is the time at which the curvature mode $k_3$ satisfies $k_3/a=m$.

Using the formula (\ref{resonance_formula}), we get
\bea
\sim c_- \left( \frac{m}{|H_{k_3}|} \right)^{1/2} \left(\frac{2k_1}{k_3} \right)^{-\frac{3}{2}+\frac{1}{2p}}
\exp\left[ i \frac{p^2}{1-p} \frac{m}{H_{k_3}} \left( \frac{2k_1}{k_3} \right)^{1/p} \mp i\frac{\pi}{4} \right] ~.
\label{Clock_signal_temp}
\eea
In general, the coefficient $c_-$ depends on $k_3$, and this introduces scenario- and model-dependent functions of $k_3$ in both the prefactor and the phase. The parameters $H_{k_3}$ and $k_3$ are evaluated at the reference point $t_3$ at which $k_3/a=m$, and this reference point may be changed arbitrarily because $H_k k^{1/p}$ remains a constant. For convenience we choose it to be $H_{k_3}$ and $k_3$. The term (\ref{Clock_signal_temp}) can then be written as
\bea
\sim f(k_3)~ \left( \frac{2k_1}{k_3} \right)^{-\frac{3}{2}+\frac{1}{2p}}
\exp\left[ i \frac{p^2}{1-p} \frac{m}{H_{k_3}} \left( \frac{2k_1}{k_3} \right)^{1/p} + i\varphi(k_3) \right] ~.
\label{Clock_signal_1}
\eea
Note that $f$ and $\varphi$ may also depend on $p$ and $m/H_{k_3}$.

We can now see the reasons of our earlier comments that the BD state origin of either the curvature scalar field or the clock field is not essential for the clock signal.
For the curvature scalar field, the non-BD component can also resonate with the clock field, similar to \eqref{clock_integral}, although the component of the clock field that it resonates with is different.
For the clock field, its non-BD state affects the detailed $k_3$-dependence in $c_-$, and is reflected as part of the arbitrariness of $f(k_3)$ and $\varphi(k_3)$ in \eqref{Clock_signal_1}, which is not the essential part of the clock signal.

The integrals in \eqref{Term2} are more difficult to be carried out, and it will be one of the main goals of the next two sections where we work out these integrals in both the inflation and alternative-to-inflation scenarios. It turns out that, due to the classical-like nature of the clock field, the final result of the clock signal in the squeezed limit is proportional to \eqref{Clock_signal_1}.

As a summary, the overall clock signal manifests as a shape-dependent oscillatory component in the squeezed limit of the three-point function. To standardize the convention, in this paper we will quote it as a component $S^{\rm clock}$ of the shape function $S$ defined in the following way \cite{Chen:2010xka}
\bea
\langle \zeta^3 \rangle \equiv S(k_1,k_2,k_3) \frac{1}{(k_1k_2k_3)^2} \tilde P_\zeta^2 (2\pi)^7 \delta^3(\sum_{i=1}^3 \bk_k) ~,
\label{S_def}
\eea
where $\tilde P_\zeta \equiv 2.1\times 10^{-9}$ is the fiducial power spectrum. In this convention, the clock signal in the squeezed limit $k_3/k_{1,2}\to 0$ is proportional to
\bea
S^{\rm clock}
&\propto
\high{
\left( \frac{2k_1}{k_3} \right)^{-\frac{1}{2} + \frac{1}{2p}}
}
\sin \left[ \frac{p^2}{1-p} \frac{m}{H_{k_3}} \left( \frac{2k_1}{k_3} \right)^{1/p} + \varphi(k_3) \right]
\label{Clock_singals_1}
\\
&\propto
\high{
\left( \frac{2k_1}{k_3} \right)^{-\frac{1}{2} + \frac{1}{2p}}
}
\sin \left[ p~ \frac{m}{m_{{\rm h},k_3}} \left( \frac{2k_1}{k_3} \right)^{1/p} + \varphi(k_3) \right] ~.
\label{Clock_signals}
\eea
In the second line we have converted the Hubble parameter $H$ to the horizon mass scale $\mhor$ defined in \eqref{mhor_def}. For non-exponential-inflation models these two parameters are different, and the horizon mass scale is more relevant physically. The parameter $m_{{\rm h},k_3}$ means that it is the value of $\mhor$ evaluated when $k_3/a=m$.

We caution that, in quoting the above result from \eqref{Term1} and \eqref{Term2}, we have assumed the prefactors $u_{k_i}|_{\tau=0}$ in those formulae to be proportional to $1/\sqrt{k_i^3}$. This is true for inflation and matter contraction; but may be non-trivial to realize for other scenarios. In any case this is only a convention.

As we can see from (\ref{Clock_signals}), the most important clock signal is the oscillatory pattern
\bea
\sin \left[ p~ \frac{m}{m_{{\rm h},k_3}} \left( \frac{2k_1}{k_3} \right)^{1/p} + \varphi(k_3) \right] ~.
\eea
Its envelop behavior also has some overall characteristic trend for each scenario, but may be subject to deformation due to model-building complexities.\footnote{\label{Footnote:Boltzmann_envelop} There is another important envelop factor that we ignore here, which is the analogous Boltzmann suppression factor that describes the probability of quantum mechanically creating a massive particle from the time-dependent background. For inflation this factor is of order $\sim \exp(-\pi m/H)$ \cite{Arkani-Hamed:2015bza}. Because $H$ is approximately constant for inflation, this factor only gives an overall suppression to the amplitude, but not any $k_1/k_3$-dependent modulation. This is no longer true for the alternative-to-inflation scenarios. In these scenarios, we expect some analogous suppression factors too if the clock field oscillation has the quantum origin, with $H$ replaced by $\mhor$. However, since $\mhor$ is no longer a constant, these factors are expected to give some additional envelop behavior to the clock signal. These modulations are only negligible if we look at very short ranges of $k_1/k_3$.\\ \indent We also comment that for some of the alternative scenarios, classical origin of the clock field is also possible, and in this case the amplitude may not be exponentially suppressed. \\ \indent
We leave these interesting issues for future investigations.}
This clock signal may be extracted from data if we fix the long-mode $k_3$ and examine the bispectrum as a function of the short mode $k_1$. This dependence gives an oscillatory signal that is similar to that of the classical standard clock \cite{Chen:2011zf,Chen:2014cwa}. However, while the clock signal is a function of the overall scale in the classical standard clock case, here it is a function of the shape of bispectrum. These clock signals directly encode the evolution of $a(t)\sim t^p$.
We also note that, with fixed $k_3$, for the expansion scenarios, the modes with smaller $k_1$ resonate first; for the contraction scenarios, the modes with larger $k_1$ resonate first.

\begin{figure}[t]
  \centering
  \includegraphics[width=0.9\textwidth]{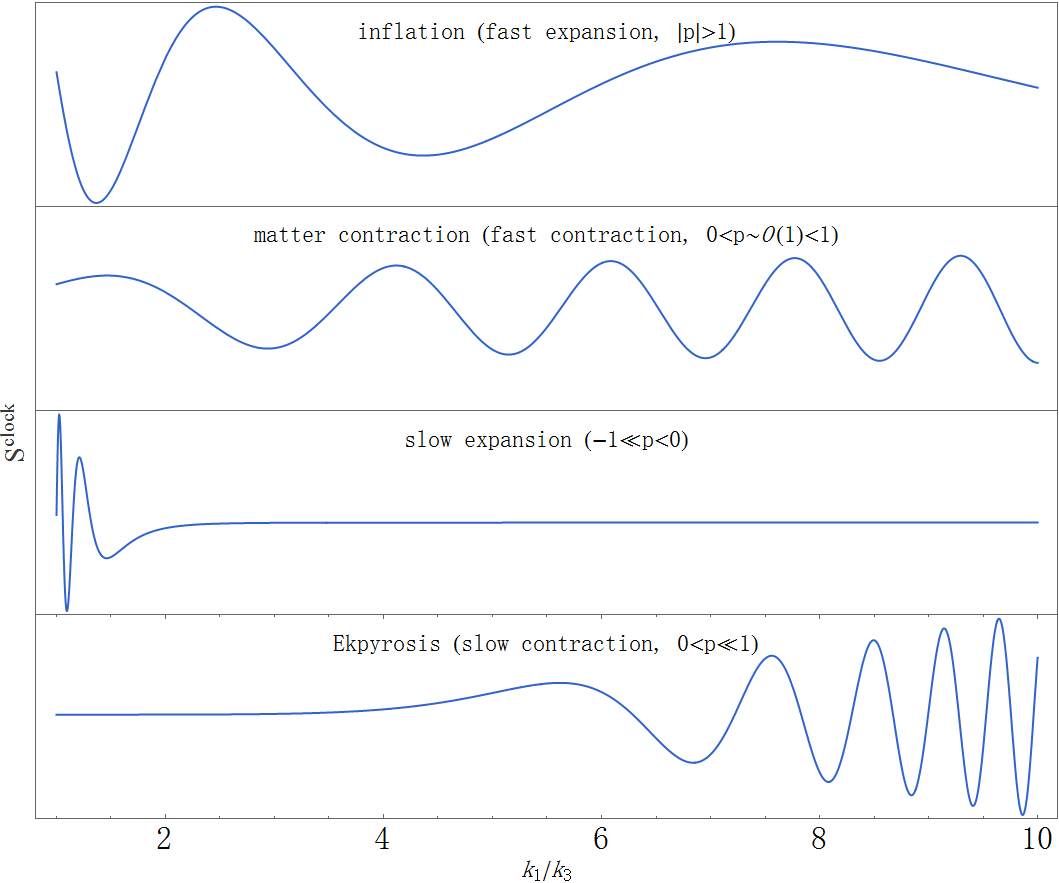}
\caption{\small Examples of clock signals in four different scenarios: inflation ($p=20$, $m/m_{{\rm h},k_3}=5$), fast contraction ($p=2/3$, $m/m_{{\rm h},k_3}=0.5$), slow expansion ($p=-0.2$, $m/m_{{\rm h},k_3}=2000$), and slow contraction ($p=0.2$, $m/m_{{\rm h},k_3}=5\times 10^{-5}$). Also, see the last paragraph of Sec.~\ref{Sec:QuantumSC} and footnote \ref{Footnote:Boltzmann_envelop} for some comments on this figure.}
\label{Fig:compare_scenarios}
\end{figure}

In Fig.~\ref{Fig:compare_scenarios}, for illustration and comparison, we sketch examples of clock signals \eqref{Clock_signals} in four different scenarios. We have chosen the parameters $m/m_{{\rm h},k_3}$ so that the four examples have similar oscillatory resolution. The following is some comments on the values of $m/m_{h,k_3}$ used in this figure. Although for inflation $\mhor$ is constant, for alternative scenarios it is not and so becomes $k$-dependent. Even if we choose the value of $m/m_{{\rm h},k_3}$ to be less than one for some scenarios in this figure, it does not mean that the clock mass is less than the horizon mass when the clock signal is generated. For example, for the slow-contraction scenario, longer modes resonate later and the scale factor $a$ evolves very slowly; so the condition $k_3/a=m$ is satisfied long after the clock signal is generated for the short mode $k_1$. Since the horizon size shrinks fast in this scenario, $m/\mhor$ has dropped significantly to $m/m_{{\rm h},k_3}$, even though it is larger than one when the clock signal for the short mode $k_1$ is generated.
That is why in Fig.~\ref{Fig:compare_scenarios}, the value of $m/m_{{\rm h},k_3}$ is so small for this scenario.
Vice versa for the slow-expansion scenario.

\section{The inflation scenario}
\label{Sec:Inflation}
\setcounter{equation}{0}

In this section, we show the details of the three-point correlation function in Fig.~\ref{Fig:3pt_vertex} in a single field inflation model with a clock field, which is essentially a quasi-single field inflation model. We pay special attention to the clock signal in this correlation function.

Note that the purpose of this section is not meant to present a complete answer for a model. Instead, it only provides one example among several generic coupling terms in the quasi-single field inflation models \cite{Chen:2009we,Baumann:2011nk,Noumi:2012vr,Arkani-Hamed:2015bza}.
Some of the results in this calculation have been obtained in \cite{Chen:2009we} for $m<3H/2$ and in \cite{Arkani-Hamed:2015bza,Noumi:2012vr} for $m>3H/2$. Here we provide more complete and complementary results for this example.
While most of the intermediate results apply to arbitrary $m$, we will be most interested in the mass range $m>3H/2$ in the final results.

For inflation, the equations of motion for the massless curvature scalar $u_k$ and the massive clock field $v_k$ are, respectively,
\bal
u_k'' - \frac{2}{\tau} u_k' + k^2 u_k =0 ~,
\\
v_k'' - \frac{2}{\tau} v_k' + k^2 v_k + \frac{m^2}{H^2\tau^2} v_k =0~.
\eal
Choosing the BD vacuum for both fields and normalizing the solutions using the canonical quantization conditions, we get
\bal
u_k &= \frac{H}{2\sqrt{\epsilon}\mpl} \frac{1}{k^{3/2}} (1+ik\tau) e^{-ik\tau} ~,
\nonumber \\
v_k &= -ie^{-\frac{\pi}{2}\mu + i \frac{\pi}{4}} \frac{\sqrt{\pi}}{2} H
(-\tau)^{3/2} H^{(1)}_{i\mu}(-k\tau) ~,
\label{mode_functions_inflation}
\eal
where
\bea
\mu = \sqrt{\frac{m^2}{H^2}- \frac{9}{4}} ~.
\label{mu_def}
\eea
In the definition of the shape function $S$ \eqref{S_def}, the formula for the power spectrum in this case is $P_\zeta = H^2/(8\pi^2 \epsilon \mpl^2)$.

Since the single field inflation model is a well-established model, we are able to use the full mode functions \eqref{mode_functions_inflation} without having to use various approximations in Sec.~\ref{Sec:Quantum_massive} and Sec.~\ref{Sec:QSC}. Some qualitative properties that we demonstrated in Sec.~\ref{Sec:Quantum_massive} and Sec.~\ref{Sec:QSC}, such as the appearance of the classical regime and the functional form of the clock signals, will be borne out in the rigorous calculations. We will also see some additional properties that are not so obvious from the qualitative arguments, such as the appearance of the Boltzmann suppression factor and the clock signal from the non-resonance integral.

We would like to compute two types of integrals -- \eqref{Term1} which only involves single layer integrals without time-ordering, and \eqref{Term2} which includes two-layer time-ordered integrals.
A summary is provided at the end of this section.

\subsection{The non-time-ordered integral}
\label{Sec:non-T-ordered}

Plugging in the mode functions \eqref{mode_functions_inflation},
both integrals in \eqref{Term1} can be done exactly. The result is
\bal
S_{\eqref{Term1}} =
\frac{c_2c_3}{\epsilon H \mpl^2}
(k_1k_2k_3)^{1/2} {\cal I}_1 + {\rm c.c.} + {\rm 2~perm.} ~,
\label{S_term1_full}
\eal
where
\bal
& {\cal I}_1 \equiv
-\frac{\sqrt{\pi}}{16 \sqrt{2}}
\frac{{\rm sech}(\pi\mu)}{\Gamma(1-i\mu)}
\frac{(k_1k_2)^{1/2}}{(k_1+k_2)^{5/2}}
\nonumber \\
\times
& \left\{ -\left[\frac{2(k_1+k_2)}{k_3}\right]^{-i\mu} \Gamma(\frac{5}{2}+i\mu) \Gamma(1-i\mu)
\Gamma(-i\mu)
~{}_2F_1\left( \frac{5}{4}+\frac{i\mu}{2},\frac{7}{4}+\frac{i\mu}{2},1+i\mu, \left(\frac{k_1+k_2}{k_3}\right)^{-2} \right)
\right.
\nonumber \\
& \left.
+ i\pi \left[\frac{2(k_1+k_2)}{k_3}\right]^{i\mu} {\rm csch}(\pi\mu) \Gamma(\frac{5}{2}-i\mu)
~{}_2F_1\left( \frac{5}{4}-\frac{i\mu}{2},\frac{7}{4}-\frac{i\mu}{2},1-i\mu, \left(\frac{k_1+k_2}{k_3}\right)^{-2} \right)
\right\} ~.
\label{I_1-exact}
\eal
In \eqref{S_term1_full}, ``2 perm" stands for two terms with the momentum permutations $k_1 \leftrightarrow k_3$ and $k_2 \leftrightarrow k_3$, respectively.

We can already see the shape-dependent oscillation factors $[2(k_1+k_2)/k_3]^{\pm i\mu}$ as long as $\mu$ is positive.
The clock signal is cleaner if we look at the limit in which the momentum triangle configuration is {\em very} squeezed, $k_3/k_1 \ll 1/\sqrt{\mu}$, and hence the hypergeometric functions in \eqref{I_1-exact} can be approximated as 1,
\bal
{\cal I}_1
\xrightarrow{\rm very~squeezed}
&
-\frac{\sqrt{\pi}}{2^7}
{\rm sech}(\pi\mu)~ k_1^{-3/2}
\nonumber \\
\times
&
\left[
i\pi
\frac{{\rm csch}(\pi\mu) \Gamma(\frac{5}{2}-i\mu)}{\Gamma(1-i\mu)}
\left( \frac{4k_1}{k_3} \right)^{i\mu}
-
\Gamma(\frac{5}{2}+i\mu) \Gamma(-i\mu)
\left( \frac{4k_1}{k_3} \right)^{-i\mu}
\right]
\nonumber \\
+&
 \CO\left[ k_1^{-3/2} \left( \frac{k_1}{k_3} \right)^{-2\pm i\mu} \right]
~.
\label{Term1_squeezed}
\eal
Notice that, in the limit $k_3/k_1\ll 1/\sqrt{\mu}$, the two momentum-permutation terms in \eqref{S_term1_full} no longer contribute to the clock signal.

Observationally we would be most interested in the region where the mass of the clock field is comparable to the Hubble parameter $H$.
If the mass becomes much larger, the quantum creation of the massive field should be Boltzmann-suppressed \cite{Arkani-Hamed:2015bza}.
To see this Boltzmann suppression factor explicitly, we can further take the large mass limit $\mu \gg 1$,
\bal
S^{\rm clock}_{\eqref{Term1}}
\xrightarrow[\rm large~mass]{\rm very~squeezed}
&
e^{i\frac{\pi}{4}}
\frac{\pi^{3/2}}{32} \frac{c_2c_3}{\epsilon H \mpl^2}
\mu^{3/2} e^{-2\pi\mu}
\left( \frac{k_1}{k_3} \right)^{-1/2}
\left[ -\left( \frac{4k_1}{k_3} \right)^{-i\mu} + i\left( \frac{4k_1}{k_3} \right)^{i\mu} \right]
+ {\rm c.c.}
\label{Term1_large_mu_squeezed_1}
\\
=&
-\frac{\pi^{3/2}}{8} \frac{c_2c_3}{\epsilon H \mpl^2}
\mu^{3/2} e^{-2\pi\mu}
\left( \frac{k_1}{k_3} \right)^{-1/2}
\sin \left[
\mu \ln \frac{4k_1}{k_3} + \frac{\pi}{4}
\right] ~.
\label{Term1_large_mu_squeezed}
\eal

As explained in Sec.~\ref{Sec:QSC}, this clock signal is a result of the resonance effect between the BD vacuum behavior of the inflaton $e^{-ik\tau}$ and the clock field oscillation $e^{i m t}$ in the classical regime. From the exact mode function \eqref{mode_functions_inflation}, we can see that this classical regime appears as the following late-time expansion,
\bea
v_k \to -i e^{-\frac{\pi}{2}\mu + i\frac{\pi}{4}} \frac{\sqrt{\pi}}{2}
H (-\tau)^{3/2}
\left[
\frac{1+{\rm coth}(\pi \mu)}{\Gamma(i\mu+1)} \left( \frac{-k\tau}{2} \right)^{i\mu}
-i \frac{\Gamma(i\mu)}{\pi} \left( \frac{-k\tau}{2} \right)^{-i\mu}
\right] ~.
\label{vk_expansion}
\eea
Strictly speaking this leading term behavior is valid for $k\tau \ll -\sqrt{\mu}$. However, for $k\tau \ll -\mu$ (i.e.~$k/a \ll m$) the characteristic oscillations necessary for the clock field are already present, although the oscillation phase is not precisely described by \eqref{vk_expansion}.

We note that such a resonance only contributes to the scaling power $-i\mu$ of the momentum ratio, while in \eqref{Term1_large_mu_squeezed_1} we have $i\mu$ as well. This comes from a non-resonance contribution of the clock field contributed by the 1st term in \eqref{vk_expansion}, and turns out to be a special case of this single layer integral in the inflation model. We elaborate this in more details in Appendix \ref{App:resonance_inflation}.

\subsection{The time-ordered integral}
\label{Sec:time-ordered_int}

To calculate \eqref{Term2}, it is convenient to switch the inner and outer integrand of the first integral and rewrite \eqref{Term2} as
\begin{align}
   2 \int_{-\infty}^{0} d\tau_1 a^2 v_{k_3}^* u'^{*}_{k_1} u'^{*}_{k_2}
    \int_{\tau_1}^0 d\tau_2 a^3 v_{k_3} u'^*_{k_3}
  +2 \int_{-\infty}^0 d\tau_1 a^2 v_{k_3} u'^{*}_{k_1} u'^{*}_{k_2}
    \int_{-\infty}^{\tau_1} d\tau_2 a^3 v^*_{k_3} u'^*_{k_3}
    + {\rm c.c.} + {\rm 2~perm.}
    ~.
\label{Term2_rewrite}
\end{align}
Then both $\tau_2$ integrals can be calculated analytically. Inserting the result of the $\tau_2$ integrals, the three-point function now contains one integral over $\tau_1$,
\bal
S_{\eqref{Term2}} = \frac{c_2c_3}{\epsilon H\mpl^2} (k_1k_2k_3)^{1/2} \mathcal{I}_2 + {\rm c.c.} + {\rm 2~perm.} ~,
\label{S_term2_result}
\eal
where
\begin{align}
  \mathcal{I}_2 \equiv \mathcal{I}_\mathrm{Hankel} + \mathcal{I}_\mathrm{2F2}
  = \int_{-\infty}^{0} d\tau
  \left(\mathcal{F}_\mathrm{Hankel} + \mathcal{F}_\mathrm{2F2}\right) ~.
\label{I2_def}
\end{align}
$\mathcal{F}_\mathrm{Hankel}$ and $\mathcal{F}_\mathrm{2F2}$ are defined as
\begin{align}
  \mathcal{F}_\mathrm{Hankel} \equiv
  -\frac{(1+i)\pi^{3/2}}{16 \cosh(\pi\mu)} e^{-\frac{\pi\mu}{2}} e^{i(k_1+k_2)\tau}
  \tau \sqrt{-k_1k_2\tau} H_{i\mu}^{(1)}(-k_3 \tau)~,
\end{align}
\begin{align}
  \mathcal{F}_\mathrm{2F2} \equiv
-\frac{2^{-3-i \mu } \sqrt{k_1 k_2 k_3} \tau ^2 \Gamma (-i \mu ) e^{-\pi  \mu +i
   \left(k_1+k_2\right) \tau } \left(-\tau  k_3\right){}^{i \mu } \left(H_{i \mu
   }^{(1)}\left(-\tau  k_3\right)+e^{\pi  \mu } H_{-i \mu }^{(2)}\left(-\tau  k_3\right)\right) \,
   _2{\hat F}_2}{2 \mu -i}
   \nonumber \\
   +\frac{2^{-3+i \mu } \sqrt{k_1 k_2 k_3} \tau ^2 \Gamma (i \mu )
   e^{-\pi  \mu +i \left(k_1+k_2\right) \tau } \left(-\tau  k_3\right){}^{-i \mu } \left(e^{\pi
   \mu } H_{i \mu }^{(1)}\left(-\tau  k_3\right)+H_{-i \mu }^{(2)}\left(-\tau  k_3\right)\right) \,
   _2\tilde{F}_2}{2 \mu +i}~,
\end{align}
where
\bal
{}_2\hat{F}_2 &\equiv {}_2F_2\left(i \mu +\frac{1}{2},i \mu +\frac{1}{2};i \mu +\frac{3}{2},2 i \mu +1;2 i \tau k_3\right)~,
\nonumber \\
{}_2\tilde{F}_2 &\equiv {}_2F_2\left(\frac{1}{2}-i \mu ,\frac{1}{2}-i \mu ;\frac{3}{2}-i \mu ,1-2 i \mu ;2 i \tau k_3\right)~.
\eal
The $\mathcal{F}_\mathrm{Hankel}$ term comes from the 2nd time-ordered integral in \eqref{Term2_rewrite}: the inner integral of this double integral has a constant part which is then turned to the $\mathcal{F}_\mathrm{Hankel}$ term.

In the remainder of this subsection, we shall show that $\mathcal{I}_\mathrm{Hankel}$ can be calculated exactly, and has clock signal in it. The $\mathcal{I}_\mathrm{2F2}$ can be calculated approximately overall, but exactly in some special limits; and it does not contain clock signal.

{\em $\bullet$ The $\mathcal{I}_\mathrm{Hankel}$ part.}
The $\mathcal{I}_\mathrm{Hankel}$ part can be integrated directly:
\begin{align} \label{eq:hankel-exact}
  \mathcal{I}_\mathrm{Hankel} = &
-\frac{ \sqrt{\pi}
   ~\text{sech}(\pi  \mu )
   }{ 8\sqrt{2} ~\Gamma (i \mu +1)}
   \frac{(k_1k_2)^{1/2}}{ (k_1+k_2)^{5/2} }
   \nonumber\\ &
   \times \left\{
   \left[ \frac{2(k_1+k_2)}{k_3} \right]^{-i \mu}
   \Gamma \left( \frac{5}{2} + i \mu \right)
   \left[\pi -i \cosh (\pi  \mu ) \Gamma (i \mu +1)
   \Gamma (-i \mu ) \right] \, _2\hat{F}_1
   \right.
   \nonumber\\ &
   \left.
   ~~~-i \left[ \frac{2(k_1+k_2)}{k_3} \right]^{i \mu} \Gamma \left(\frac{5}{2}-i \mu \right) \Gamma (i \mu +1) \Gamma
   (i \mu ) e^{-\pi\mu} \, _2\tilde{F}_1
   \right\} ~,
\end{align}
where
\bal
{}_2\hat{F}_1 &\equiv \, _2F_1\left( \frac{5}{4} + \frac{i \mu}{2},\frac{7}{4} + \frac{i \mu}{2}; 1+ i\mu;
\left( \frac{k_1+k_2}{k_3}\right)^{-2} \right)~,
\nonumber \\
{}_2\tilde{F}_1 &\equiv \, _2F_1\left( \frac{5}{4}-\frac{i \mu}{2}, \frac{7}{4}- \frac{i \mu}{2} ;1-i \mu ;
\left( \frac{k_1+k_2}{k_3}\right)^{-2}
   \right)~.
\eal
Again, one can already observe the oscillatory factors $\left[ 2(k_2+k_2)/k_3\right]^{\pm i\mu}$ for positive $\mu$.
The clock signal is cleaner once we expand \eqref{eq:hankel-exact} around the very squeezed limit $k_3/k_1\ll 1/\sqrt{\mu}$,
\begin{align}
\mathcal{I}_\mathrm{Hankel} \xrightarrow{\rm very~squeezed}
  - 2^{-6} \sqrt{\pi }
  k_1^{-3/2} &
   \left[
   \frac{\pi~ \Gamma \left(i \mu +\frac{5}{2}\right) (\text{csch}(\pi  \mu )+\text{sech}(\pi
   \mu ))}{\Gamma (i \mu +1)}
   \left(\frac{4k_1}{k_3}\right)^{-i \mu }\right.
   \nonumber \\ &
   - \left.
   \frac{2i~ \Gamma \left(\frac{5}{2}-i \mu \right) \Gamma
   (i \mu ) }{e^{2 \pi  \mu }+1}
      \left(\frac{4k_1}{k_3}\right)^{ i \mu }
   \right]
   + \CO\left[ k_1^{-3/2} \left( \frac{k_1}{k_3} \right)^{-2\pm i\mu} \right]
   ~.
\label{eq:IHankel_squeezed}
\end{align}
Again note that, in the limit $k_3/k_1\ll 1/\sqrt{\mu}$, the two momentum-permutation terms in \eqref{S_term2_result} no longer contribute to the clock signal.

{\em $\bullet$ The $\mathcal{I}_\mathrm{2F2}$ part.}
We are unable to integrate the $\mathcal{I}_\mathrm{2F2}$ part directly. Nevertheless, one can use the method of resummation following Ref.~\cite{Chen:2012ge}.
Namely, we first Taylor-expand the hypergeometric functions in terms of $\tau$ around $\tau=0$; integrate each term in the Taylor-expansion with respect to $\tau$; and then try to resum the resulting series or approximate the sum by a finite number of terms.

Following this procedure, we find that, after integrating over $\tau$, the resulting series is in terms of powers of $k_{\rm massive}/(k_{\rm massless1}+k_{\rm massless2})$. This is also true for the other two momentum-permutation terms. So at least numerically this method can be used to compute most of the momentum configurations (except for the folded limit) to arbitrary precision, although in some cases the convergence is slow.

\begin{figure}[t]
  \centering
  \includegraphics[width=0.45\textwidth]{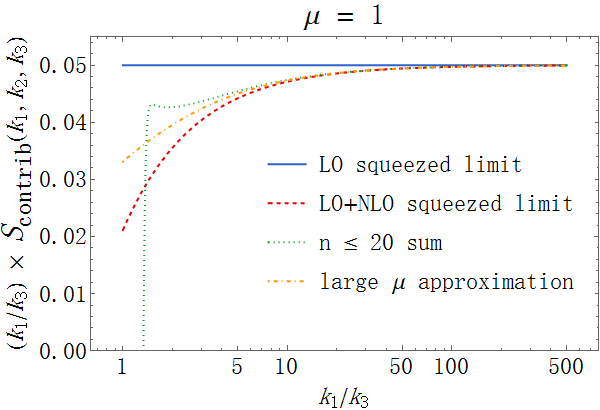}\hspace{0.05\textwidth}
  \includegraphics[width=0.45\textwidth]{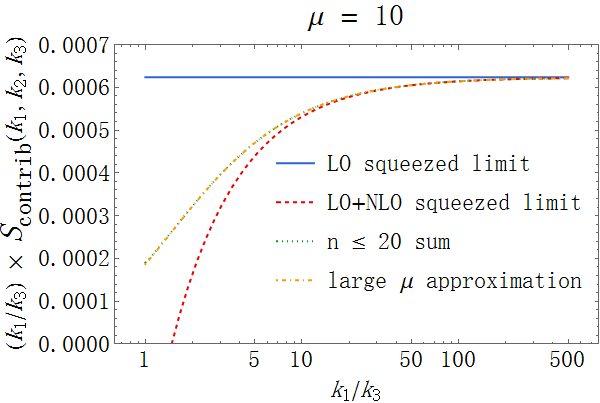}
  \caption{\label{fig:2F2-converge} \small We demonstrate that both the squeezed-limit approximation \eqref{eq:I2F2_squeezed} and the large $\mu$ limit exact result \eqref{eq:large_mu_limit} give good approximations to the full ${\cal I}_{\rm 2F2}$. The vertical axis is the corresponding contribution to the shape function, and the unit of $S_{\rm contrib}$ is $2c_2c_3/(\epsilon H \mpl^2)$. We used the isosceles configuration $k_1=k_2$. }
\end{figure}

To analytically investigate whether the $\mathcal{I}_\mathrm{2F2}$ part contains the clock signal, as explained we need to look at the case in which the massive field takes the long mode and two (of the three) massless modes are the short modes, $k_{\rm massive}\ll k_{\rm massless1,2}$. To do that we only need to look at the integral ${\cal I}_{\rm 2F2}$ in which the wavenumber of the massive field $k_3$ is the long mode; the two other momentum-permutation terms in \eqref{S_term2_result} are not needed because they correspond to the configuration in which the massive mode is a short mode. The series converge fairly well in this limit, and we find that the first two terms in the series already give fairly good approximation to the full result.  To reach the similar precision order in terms of $k_3/k_1$ as in \eqref{Term1_squeezed} and \eqref{eq:IHankel_squeezed}, we need take the squeezed limit of both terms,
\begin{align}
  \mathcal{I}_\mathrm{2F2} \xrightarrow{\rm squeezed}
   \frac{\sqrt{k_3}}{4 (4\mu^2+1) k_1^2}
   - \frac{3k_3^{3/2}}{8 (4\mu^2 + 9) k_1^3}
   + \CO\left[ k_1^{-3/2} \left( \frac{k_1}{k_3} \right)^{-5/2} \right] ~,
\label{eq:I2F2_squeezed}
\end{align}
where the first term comes from the 0th order Taylor expansion and the second term comes from the 1st order. Notice that, unlike the previous cases where the condition for the very squeezed limit is $\mu$-dependent, here the validity of \eqref{eq:I2F2_squeezed} only requires $k_3/k_1 \ll 1$, independent of $\mu$. See Fig.~\ref{fig:2F2-converge} for some examples. This is because, as mentioned, the Taylor-series from the resummation method is a series in order of $k_3/k_1$; and in the several leading order terms that we pick up from the series, the hypergeometric functions take the forms such as $_2F_1\left(2,\frac{3}{2};1\pm i \mu ; \frac{k_3^2}{4k_1^2} \right)$ and $_2F_1\left(2,\frac{5}{2};1\pm i \mu;\frac{k_3^2}{4k_1^2} \right)$. Unlike those in ${\cal I}_1$ and ${\cal I}_{\rm Hankel}$, these hypergeometric functions have only one $\mu$-dependent parameter and hence the validity condition for the small $k_3/k_1$ expansion no longer relies on the value of $\mu$.

One can see clearly that there is no clock signal in ${\cal I}_{\rm 2F2}$.

\begin{figure}[t]
  \centering
  \includegraphics[width=0.43\textwidth]{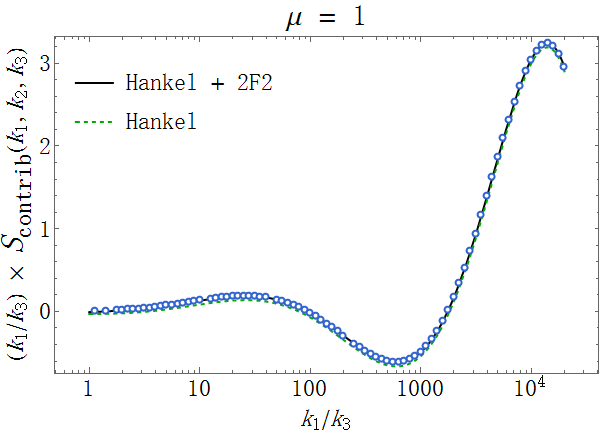}\hspace{0.05\textwidth}
  \includegraphics[width=0.465\textwidth]{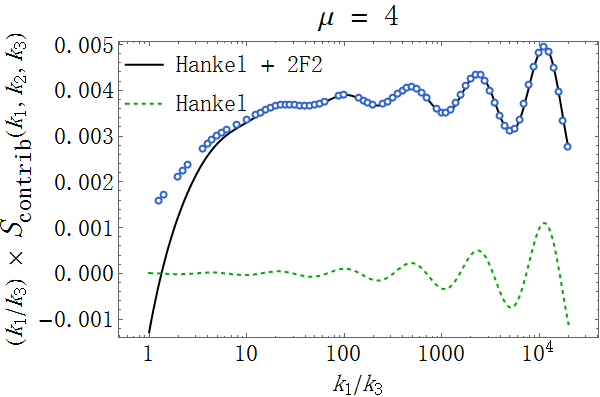}
  \caption{\label{fig:clocks} \small
  A comparison between the squeezed limit approximation, \eqref{eq:IHankel_squeezed} $+$ \eqref{eq:I2F2_squeezed}, and the full numerical calculation of the bispectrum where the massive field is the long mode. The isosceles configuration $k_1=k_2$ is used. The vertical axis is the corresponding contribution to the shape function, and the unit of $S_{\rm contrib}$ is $2c_2c_3/(\epsilon H \mpl^2)$. The black solid line denotes the full analytical result, and the green dashed line denotes the Hankel part separately. The blue dots line represents the numerical calculation. One observes that for small $\mu$, the Hankel part gives dominate contribution. For larger $\mu$, though the Hankel part still contribute all the clock signals, the ${}_2F_2$ part contributes a significant shift of the amplitude.}
\end{figure}

{\em $\bullet$ Large $\mu$ limit of $\mathcal{I}_\mathrm{2F2}$}. In the large $\mu$ limit (but not necessarily squeezed limit), closed-form exact results can be obtained using the method of resummation \cite{Chen:2012ge}. After one Taylor-expands the hypergeometric functions $\mathcal{F}_\mathrm{2F2}$ and integrates each term with respect to $\tau$, new hypergoemetric functions arise of the form ${}_2F_1(2,3/2;1\pm i\mu;k_3^2/(k_1+k_2)^2)$ or ${}_2F_1(2,5/2;1\pm i\mu;k_3^2/(k_1+k_2)^2)$. Note that those hypergeometric functions can be expanded in their series form as
\begin{equation}
  {}_2F_1(a,b;1\pm i\mu;z) = \sum_{k=0}^{\infty} \frac{(a)_k (b)_k}{(1\pm i\mu)_k} \frac{z^k}{k!}~,
\end{equation}
where $(a)_k \equiv \Gamma(a+k)/\Gamma(a)$ is the Pochhammer function. Note that the only term which does not have the $1\pm i\mu$ factor in denominator is the $k=0$ term. Thus in the large $\mu$ limit only the $k=0$ term contributes. The result is simply
\begin{equation}
  {}_2F_1(a,b;1\pm i\mu;z) \xrightarrow{\mu\rightarrow\infty} 1~.
\end{equation}
This greatly simplifies the calculation, and the series can then be resummed. We get
\begin{equation}
\mathcal{I}_\mathrm{2F2} \xrightarrow{\mu\rightarrow\infty}
-\frac{\sqrt{k_1 k_2 k_3} \left(\frac{\left(3 f_2+f_3\right) k_3}{2 \mu +3 i}-\frac{2 f_1 \left(k_1+k_2\right)}{2 \mu +i}\right)}
{4 \left(k_1+k_2\right){}^4 \mu } + {\rm c.c.} ~,
\label{eq:large_mu_limit}
\end{equation}
where
\begin{equation}
  f_1 \equiv \, _3F_2\left(1,\frac{1}{2}-i \mu ,\frac{1}{2}-i \mu ;\frac{3}{2}-i \mu ,1-2 i \mu ;-\frac{2 k_3}{k_1+k_2}\right)~,
\end{equation}
\begin{equation}
  f_2 \equiv \, _3F_2\left(2,\frac{3}{2}-i \mu ,\frac{3}{2}-i \mu ;\frac{5}{2}-i \mu ,2-2 i \mu ;-\frac{2 k_3}{k_1+k_2}\right)~,
\end{equation}
\begin{equation}
  f_3 \equiv \, _4F_3\left(2,2,\frac{3}{2}-i \mu ,\frac{3}{2}-i \mu ;1,\frac{5}{2}-i \mu ,2-2 i \mu ;-\frac{2 k_3}{k_1+k_2}\right)~.
\end{equation}
Note that the ``c.c." in \eqref{eq:large_mu_limit} is in addition to that in \eqref{S_term2_result}.

We also compare this large $\mu$ expression with the results with finite $\mu$ in Fig.~\ref{fig:2F2-converge}. One can see that it is a good approximation even at the smaller $\mu$ values.

{\em $\bullet$ $\mathcal{I}_\mathrm{Hankel}+\mathcal{I}_\mathrm{2F2}$.}
The following are a few aspects of the net result $\mathcal{I}_\mathrm{Hankel}+\mathcal{I}_\mathrm{2F2}$.

Comparisons between the squeezed limit approximation, \eqref{eq:IHankel_squeezed} $+$ \eqref{eq:I2F2_squeezed}, and the full numerical calculation where the massive field is the long mode are shown in Fig.~\ref{fig:clocks}.

Parallel to what we did in \eqref{Term1_large_mu_squeezed}, we can further take the large mass limit of these results.
The clock signal is again given by
\eqref{eq:IHankel_squeezed} in these limits,
\bea
S^{\rm clock}_{\eqref{Term2}}
\xrightarrow[\rm large~mass]{\rm very~squeezed}&
\high{
-\frac{\pi^{3/2}}{32} e^{i\frac{3\pi}{4}}
\frac{c_2c_3}{\epsilon H \mpl^2} \mu^{3/2} e^{-\pi\mu}
\left( \frac{k_1}{k_3} \right)^{-1/2}
\left( \frac{4k_1}{k_3} \right)^{-i\mu}
}
+ {\rm c.c.}
\\
=&
\high{
\frac{\pi^{3/2}}{16}
\frac{c_2c_3}{\epsilon H \mpl^2} \mu^{3/2} e^{-\pi\mu}
\left( \frac{k_1}{k_3} \right)^{-1/2}
}
\sin \left[
\mu \ln \frac{4k_1}{k_3} + \frac{3\pi}{4}
\right]
~,
\label{Term2_large_mu_squeezed}
\eea
where we can see the Boltzmann suppression factor.

Note that in these limits, the contribution that does not contain the clock signal, namely \eqref{eq:I2F2_squeezed}, is not suppressed by the Boltzmann factor but only by the power law $\sim \mu^{-2}$. The physical interpretation \cite{Arkani-Hamed:2015bza} is that such a contribution does not describe the quantum creation of massive particles from the inflationary background, and is captured in the low energy effective theory. However the clock signal corresponds to the quantum creation of massive particles and hence has to be suppressed by a Boltzmann factor; and this contribution is not captured in the low energy effective theory language.

\subsection{Summary}

To summarize, in this section we studied the quantum standard clock signal in the inflation model. We have presented the analytical and numerical results on the bispectrum term Fig.~\ref{Fig:3pt_vertex} for a single field inflation model with a clock field.
We are most interested in the clock signals in which there are shape-dependent oscillations.

For the clock signal, we are able to obtain the exact analytical result. In terms of the shape function $S$ \eqref{S_def}, this is given by
\bal
S^{\rm clock} =
\frac{c_2c_3}{\epsilon H \mpl^2}
(k_1k_2k_3)^{1/2}
\left( {\cal I}_1 + {\cal I}_{\rm Hankel} \right)
+ {\rm c.c.} + {\rm 2~perm.} ~,
\label{S_full_clock}
\eal
where ${\cal I}_1$ and ${\cal I}_{\rm Hankel}$ are given by \eqref{I_1-exact} and \eqref{eq:hankel-exact}, respectively.

In the very squeezed limit, $k_3/k_1 \ll 1/\sqrt{\mu}$, the appearance of the clock signal becomes cleaner, and they are given by \eqref{Term1_squeezed}+\eqref{eq:IHankel_squeezed}. Note that the other two momentum-permutation terms in \eqref{S_full_clock} do not contribute to the clock signal in this limit.

Further taking the large mass $m\gg H$ limit, we can see the Boltzmann suppression factor. The result is \eqref{Term1_large_mu_squeezed}+\eqref{Term2_large_mu_squeezed}.
Comparing \eqref{Term1_large_mu_squeezed} and \eqref{Term2_large_mu_squeezed}, we can see that \eqref{Term2_large_mu_squeezed} is the leading behavior in this limit.

As we can see explicitly in, for example, \eqref{Term1_large_mu_squeezed} and \eqref{Term2_large_mu_squeezed}, these clock signals are the special case of the general clock signal \eqref{Clock_signals} in the $p\to \infty$ limit. As expected, the function of the momentum ratio in the phase of this oscillatory signal is the inverse function of the inflationary scale factor $a\sim e^{Ht}$.

There is also another contribution ${\cal I}_{\rm 2F2}$, which we show does not contain any clock signal. For this contribution, in general we can only obtain a series expansion but not a full closed-form result. Nonetheless, the analytical result in the squeezed approximation ($k_3/k_1 \ll 1$) in which the massive field takes the long mode is obtained in \eqref{eq:I2F2_squeezed}; and the closed-form analytical result in the large $\mu$, but not necessarily squeezed, limit is obtained in \eqref{eq:large_mu_limit} using the method of resummation.

\begin{figure}[t]
  \centering
  \includegraphics[width=0.9\textwidth]{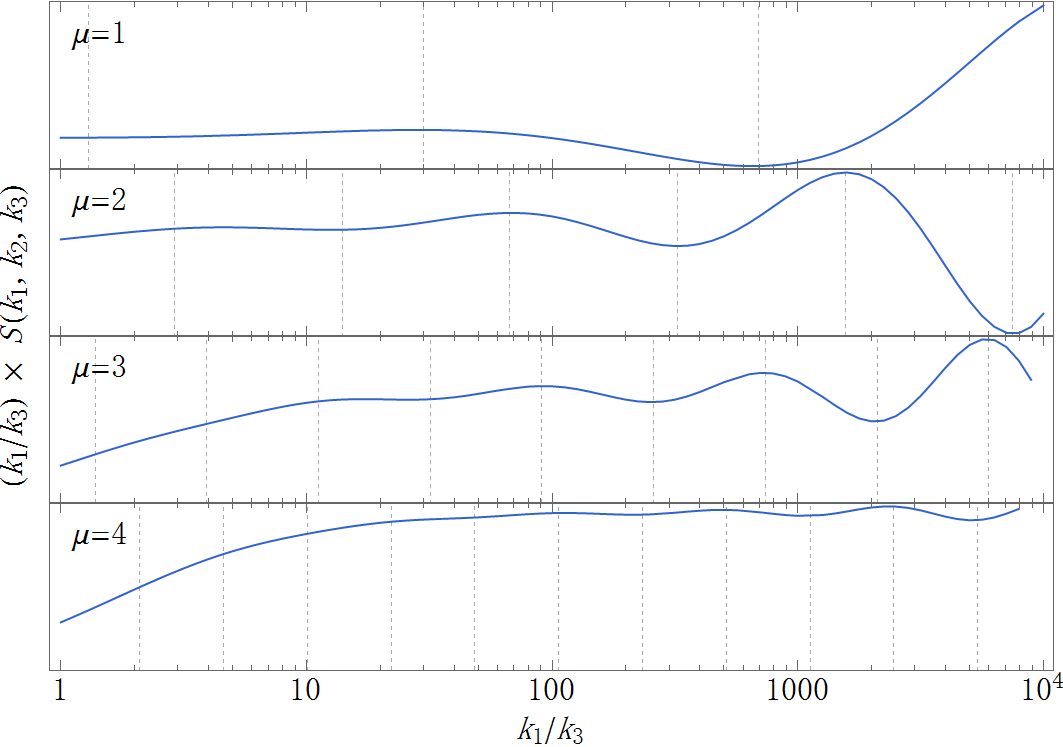}
  \caption{\label{fig:num_and_app} \small The numerical results of the full shape functions with $\mu$ of order $\CO(1)$ in the inflation scenario. These numerical results include both the clock signals and the non-clock signals. All momentum permutations are included in these results. The method of Wick rotation \cite{Chen:2009we} is applied in the numerical integrals to increase the convergence speed in the UV. To show more clearly the oscillation patterns of the clock signals, we have multiplied $k_1/k_3$ to the shape function. The vertical gray dashed lines denote the peaks of the oscillation. The periodicity in $\log (k_1/k_3)$ is $1/\mu$. This agrees with our analytical result.}
\end{figure}

In Fig.~\ref{fig:num_and_app}, we show several examples of the complete numerical results of the shape functions. These shape functions contain both the clock signals and non-clock signals. We choose $\mu$ to be of order one in this figure.

\section{The alternative scenarios ($0<p<1$)}
\label{Sec:Alternatives}
\setcounter{equation}{0}

In this section, we study the clock signals generated in the alternative-to-inflation scenarios with $0<p<1$. This includes the fast-contraction scenario $0<p\sim\CO(1)<1$, such as the matter contraction scenario \cite{Wands:1998yp}, and the slow-contraction scenario $0<p\ll 1$, such as the Ekpyrosis scenario \cite{Khoury:2001wf}. As the inflation example, we assume one massless field that is responsible for the curvature perturbation and one massive field as the clock field.

However, comments are in order before we proceed to the more detailed analyses.

As we know, alternative scenarios are not well established as inflation. For example,
in alternative scenarios, the assumption of single massless field typically does not lead to viable models; even with one massless mode, its superhorizon behavior may be highly model-dependent.

Fortunately, the standard clock signals are largely insensitive to such details. They manifest as some fast-oscillating signals added on top of the density perturbations with otherwise much smoother scale or shape dependence.
Equally importantly, in the regime where the clock signals are generated, mode functions exhibit some universal behaviors. This makes the general analyses possible. Therefore, although unlike the inflation case where we have the complete mode functions, in this section we can still proceed by utilizing these universal behaviors. We start by briefly reviewing these behaviors in the following.

For arbitrary $p$, the scale factor and the relation between $t$ and $\tau$ are as follows,
\bea
a(t) = a_0 \left( \frac{t}{t_0} \right)^p
= a_0 \left( \frac{\tau}{\tau_0} \right)^{\frac{p}{1-p}} ~,
~~~
\tau = \frac{1}{1-p} \frac{t_0^p}{a_0} t^{1-p} ~.
\eea
As we showed, the clock signal is generated when two curvature scalars are the short modes and at the sub-horizon region, and the clock field is the long mode and enters the classical regime.
We thus approximate the two short curvature scalars using the universal sub-horizon behavior,
\bea
u_k \to \frac{1}{a\sqrt{4\epsilon k}} e^{-ik\tau} ~,
\label{uk_BD}
\eea
where we have again chosen the BD vacuum;
and we approximate the clock field fluctuation $\delta\sigma$ by the solution in the classical regime \eqref{classical_limit},
\bea
v_k \propto \left( \frac{\tau_0}{\tau} \right)^{\frac{3p}{2(1-p)}}
\left[
c_+ e^{ i(1-p) \nu \left( \frac{\tau}{\tau_0} \right)^{\frac{1}{1-p}}}
+
c_- e^{-i (1-p) \nu \left( \frac{\tau}{\tau_0} \right)^{\frac{1}{1-p}}}
\right] ~.
\label{vk_alt}
\eea
In the classical regime the physical wavenumber of the clock field has effectively been taken to zero because it is much smaller than the typical frequency scale of the resonance phenomenon $\sim m$, or equivalently as we noted previously, the field is homogeneous over a region much larger than the Compton scale.
In \eqref{vk_alt}, the first component is the positive frequency mode, the second is the negative frequency mode, $\tau_0$ is a negative constant, and
\bea
\nu = m/(-a_0\tau_0)^{-1}
\label{nu_def}
\eea
is the ratio of the clock field mass to the horizon scale mass at $\tau_0$. Although the horizon mass $(-a\tau)^{-1}$ is usually time-dependent, here $\nu$ is evaluated at $\tau_0$ and by definition a constant. This ratio is an approximate generalization of $\mu$ \eqref{mu_def} defined in the inflation case.

For simplicity, we choose $\tau_0=-1$ and denote
\bal
u_k &\propto \frac{1}{(-\tau)^{\frac{p}{1-p}}} e^{-ik\tau} ~,
\nonumber \\
v_k &\propto \frac{1}{(-\tau)^{\frac{3p}{2(1-p)}}}
\left[
c_+ e^{i(1-p)\nu(-\tau)^{\frac{1}{1-p}}}
+
c_- e^{-i(1-p)\nu(-\tau)^{\frac{1}{1-p}}}
\right] ~.
\label{Modes_alternatives}
\eal
The pre-factors of $k_i$'s may be easily restored at the end of the calculation.
We will also use the above $u_k$ for the long curvature scalar mode, because for contraction scenarios the long mode is still at subhorizon during the resonance in most cases, as one can infer from Fig.~\ref{Fig:scales}.\footnote{There is also a small corner of parameter space in which the resonance happens when the long mode is at superhorizon. In such a case, $u_{k_{\rm long}} \propto (-\tau)^{-\beta}$. As one can check, as long as $\frac{p}{2(1-p)}<\beta<\frac{2+p}{2(1-p)}$, all the conclusions in this section remain the same and we get the same result \eqref{alt_double} with a different constant prefactor.
}

\subsection{The non-time-ordered integral}

The non-time-ordered integral (\ref{Term1}) is analysed in Sec.~\ref{Sec:QuantumSC}. Only the following term contains resonance in the squeezed limit $k_3/k_1 \to 0$,
\bal
\propto
c_-^*
\int^0_{-\infty} d\tau (-\tau)^{-\frac{3p}{2(1-p)}}
e^{i(1-p)\nu (-\tau)^{\frac{1}{1-p}}} e^{2ik_1\tau} ~.
\label{alt_single}
\eal
Note that only the negative energy mode $v_k$ contributes to the clock signal -- unlike the inflation case, the integrals that do not resonate generally do not contribute to the clock signals, and we demonstrate this observation in Appendix \ref{App:resonance_alter}.\footnote{As we demonstrate in Appendix \ref{App:resonance_alter}, the conclusion, that the clock signals are only contained in the resonant integrals for the alternative scenarios with $0<p<1$, are made empirically through examining several representative examples, and a general study is currently unavailable.}

\subsection{The time-ordered integral}

Using \eqref{Modes_alternatives}, the integrals (\ref{Term2}) become
\bal
\propto
&
\int^\infty_0 dx_1 x_1^{-\frac{3p}{2(1-p)}}
\left(
c_+^* e^{-i(1-p)\nu x_1^{\frac{1}{1-p}}}
+
c_-^* e^{i(1-p)\nu x_1^{\frac{1}{1-p}}}
\right)
e^{-2ik_1 x_1}
\nonumber \\
\times&
\int^{x_1}_0 dx_2 x_2^{\frac{p}{2(1-p)}}
\left(
c_+ e^{i(1-p)\nu x_2^{\frac{1}{1-p}}}
+
c_- e^{-i(1-p)\nu x_2^{\frac{1}{1-p}}}
\right)
e^{-ik_3 x_2}
\label{Term1_alt}
\\
+&
\int^\infty_0 dx_1 x_1^{-\frac{3p}{2(1-p)}}
\left(
c_+ e^{i(1-p)\nu x_1^{\frac{1}{1-p}}}
+
c_- e^{-i(1-p)\nu x_1^{\frac{1}{1-p}}}
\right)
e^{-2ik_1 x_1}
\nonumber \\
\times&
\int_{x_1}^\infty dx_2 x_2^{\frac{p}{2(1-p)}}
\left(
c_+^* e^{-i(1-p)\nu x_2^{\frac{1}{1-p}}}
+
c_-^* e^{i(1-p)\nu x_2^{\frac{1}{1-p}}}
\right)
e^{-ik_3 x_2} ~,
\label{Term2_alt}
\eal
where we have redefined $x_1=-\tau_1$ and $x_2=-\tau_2$, and in \eqref{Term1_alt} switched the inner and outer integrand for convenience.
The massive field is in the classical regime, and for the same reason we stated below \eqref{vk_alt} we can approximate $k_3=0$. The inner integrals can then be done analytically.

We first work on the integral \eqref{Term1_alt}. Integrating over $x_2$, we get
\bal
&
(-1+p) x_2^{\frac{2-p}{2(1-p)}}
\left. \left[
c_+ E_{\frac{p}{2}}\left( -i(1-p)\nu x_2^{\frac{1}{1-p}} \right)
+
c_- E_{\frac{p}{2}}\left( i(1-p)\nu x_2^{\frac{1}{1-p}} \right)
\right] \right|_0^{x_1}
\nonumber \\
\xrightarrow{\nu x_1^{\frac{1}{1-p}} \gg 1}
&
\frac{i}{\nu} x_1^{\frac{-p}{2(1-p)}}
\left[
-c_+ e^{i(1-p)\nu x_1^{\frac{1}{1-p}}}
+c_- e^{-i(1-p)\nu x_1^{\frac{1}{1-p}}}
\right]
\nonumber\\
&
-\frac{i}{\nu} \Gamma(1-\frac{p}{2})
\left[
- c_+ \left( -i(1-p) \nu \right)^{p/2}
+ c_- \left( i(1-p) \nu \right)^{p/2}
\right] ~,
\label{Term1_alt_inner}
\eal
where $E_n(z)$ is the exponential integral function. In the 2nd line, we have expanded the upper limit of integral in the large $\nu x_1^{\frac{1}{1-p}}$ limit, because the resonance happens when $m|t_1|\gg 1$ (see \eqref{Mass_cond2}), namely, when $\nu x_1^{\frac{1}{1-p}}$ is large. The lower limit of the integral gives a constant term in the third line.

Combining \eqref{Term1_alt_inner} with the integrand in the outer integral and using the observation from Appendix \ref{App:resonance_alter}, we can identify all the terms that contain resonance and hence potentially the clock signals,
\bal
&
-c_+ c_-^* \frac{i}{\nu}
\int_0^\infty dx_1 x_1^{-\frac{2p}{1-p}}
e^{2i(1-p)\nu x_1^{\frac{1}{1-p}}} e^{-2ik_1x_1}
\nonumber \\
&
+\frac{i}{\nu} \Gamma(1-\frac{p}{2})
\left[
c_+ c_-^* \left( -i(1-p) \nu \right)^{p/2}
- |c_-|^2 \left( i(1-p) \nu \right)^{p/2}
\right]
\int^\infty_0 dx_1 x_1^{-\frac{3p}{2(1-p)}}
e^{i(1-p)\nu x_1^{\frac{1}{1-p}}} e^{-2ik_1x_1} ~.
\label{alt1}
\eal
Note that the 2nd term is due to the constant term in \eqref{Term1_alt_inner}.

We now work on the second integral \eqref{Term2_alt}. Similarly, the inner integral can be done analytically and we then expand it in the large $\nu x_1^{\frac{1}{1-p}}$ limit,
\bal
&
(-1+p) x_2^{\frac{2-p}{2(1-p)}}
\left. \left[
c_+^* E_{\frac{p}{2}}\left( i(1-p)\nu x_2^{\frac{1}{1-p}} \right)
+
c_-^* E_{\frac{p}{2}}\left( -i(1-p)\nu x_2^{\frac{1}{1-p}} \right)
\right] \right|_{x_1}^\infty
\nonumber \\
\xrightarrow{\nu x_1^{\frac{1}{1-p}} \gg 1}
&
\frac{i}{\nu} x_1^{\frac{-p}{2(1-p)}}
\left[
- c_+^* e^{-i(1-p)\nu x_1^{\frac{1}{1-p}}}
+c_-^* e^{i(1-p)\nu x_1^{\frac{1}{1-p}}}
\right] ~.
\eal
Note that, at the infinity end, the inner integral gives zero because $x_2^{\frac{-p}{2(1-p)}}$ decays as $x_2\to\infty$ for $0<p<1$.\footnote{For the inflation scenario ($|p|>1$) and the slow-expansion scenario ($-1\ll p<0$), this is no longer true. These cases need to be treated separately. The inflation case is studied in Sec.~\ref{Sec:Inflation}.}
Combining with the integrand in the outer integral, we identify all terms that contain resonance as
\bal
c_+ c_-^* \frac{i}{\nu}
\int_0^\infty dx_1 x_1^{-\frac{2p}{1-p}}
e^{2i(1-p)\nu x_1^{\frac{1}{1-p}}} e^{-2ik_1x_1} ~.
\label{alt2}
\eal

It is interesting to notice that the terms with doubled clock frequency in \eqref{alt1} and \eqref{alt2} cancel each other, and the final clock signal is contained in the following term,
\bal
\frac{i}{\nu} \Gamma(1-\frac{p}{2})
\left[
c_+ c_-^* \left( -i(1-p) \nu \right)^{p/2}
- |c_-|^2 \left( i(1-p) \nu \right)^{p/2}
\right]
\int^\infty_0 dx_1 x_1^{-\frac{3p}{2(1-p)}}
e^{i(1-p)\nu x_1^{\frac{1}{1-p}}} e^{-2ik_1x_1}
~.
\label{alt_double}
\eal
This integral is of the same form as \eqref{alt_single}.

We note that both \eqref{alt_single} and \eqref{alt_double} are non-vanishing only in the presence of the negative-energy component of the clock field. This is due to our choice of the BD vacuum for the massless curvature scalar in \eqref{uk_BD}. Although the BD vacuum is the natural choice for inflation models, as discussed this does not have to be the case for contraction scenarios. In the presence of the non-BD component, the integral terms that contain resonance would be different and the positive-energy component of the clock field can also contribute.

Both integrals, \eqref{alt_single} and \eqref{alt_double}, are divergent at $\tau\to0$ and needs to be cut off or regularized through model-building.
However, as we emphasized the advantage of the standard clock is that such details do not affect the essential features of the clock signal contained in these integrals.
As in \eqref{resonance_formula}, using the saddle point approximation, we can extract the clock signal from these integrals,
\bea
S^{\rm clock} \propto (2k_1)^{-\frac{1}{2}+\frac{1}{2p}} \sin\left[ p~\nu^{1-\frac{1}{p}} (-2k_1\tau_0)^{1/p} +{\rm phase} \right] ~,
\label{clock_signal_alt}
\eea
where we have restored some pre-factors of $k_1$ and the factor of $\tau_0$.
This result matches with the general result \eqref{Clock_signals} with $0<p<1$. To see the correspondence between the constants in these two formulae, we assume that the wavenumber $k_3$ satisfies $k_3/a=m$ at $\tau=\tau_0$. We then have the following two relations, $k_3/a_0 = m$, $H_{k_3}=a'/a^2|_{\tau_0}=p/\left((1-p)a_0\tau_0\right)$. Together with \eqref{nu_def}, these relations connect the conventions in the two formulae. Finally, as we cautioned below \eqref{Clock_signals}, we note again that, when quoting the convention $S$ here, $u_{k_i}|_{\tau\to 0} \propto 1/\sqrt{k_i^3}$ has been assumed for those prefactors in \eqref{Term1} and \eqref{Term2}, which may be nontrivial to achieve for alternative-to-inflation scenarios.

It is interesting to observe that, in the time-ordered double integrals, naively there are terms in \eqref{alt1} and \eqref{alt2} containing clock signals with doubled oscillatory frequency; but they get  cancelled in the end. A similar cancellation also appears in the inflation case, where we showed that the integral ${\cal I}_{\rm 2F2}$ does not contain clock signals.
In the inflation case, this fact can be readily understood \cite{Arkani-Hamed:2015bza}: as the wavenumber of the massive field approaches zero, two vertices in Fig.~\ref{Fig:3pt_vertex} are separated by the lightcone and so the non-analytic terms in the propagator between them do not depend on the time-ordering. Thus the overall integral that contains non-analytic terms in $k$ factorizes. Since we used the exact mode functions in the inflation example, this property is automatically present through the $k$-dependence of the coefficients $c_+$ and $c_-$. But in the analyses of the alternative-to-inflation scenarios, we did not use any properties of $c_+$ and $c_-$;\footnote{Even if the complete model and mode functions were given, in the classical regime the clock field is mostly at the subhorizon scale for the contraction scenarios (see Fig.~\ref{Fig:scales}).} nonetheless we still see a result that indicates some kind of factorizability in the integral. So, we expect a different and more general explanation: when we set $k_3\to 0$ and use the behavior of the massive field in the classical regime, the massive mode at the first vertex of Fig.~\ref{Fig:3pt_vertex} acts much like the background oscillation in the case of the two-point function of the classical standard clock.

\subsection{Summary}
To summarize, in this section we studied the quantum standard clock signal in the alternative-to-inflation scenarios ($0<p<1$) including both the slow and fast contraction scenarios, again using the example Fig.~\ref{Fig:3pt_vertex}. Despite of the uncertainties and immaturity of the actual model-building for these scenarios, we explained how we can use the universal behaviors of the mode functions \eqref{uk_BD} and \eqref{vk_alt} to extract the essential features of the clock signal from the integrals. We showed that both the non-time-ordered integral \eqref{Term1} and time-ordered integral \eqref{Term2} are proportional to the same resonant integral, which gives rise to the clock signal \eqref{clock_signal_alt}. As expected, the functional form of the phase of this oscillatory signal is the inverse function of the background scale factor evolution, $a\sim t^p$.

\section{Amplitude of the signals}
\setcounter{equation}{0}
\label{Sec:Amplitude}

There are a variety of well-established inflation models and it is possible to introduce some massive fields and study them case by case. In this section, we briefly discuss the amplitude of the clock signals using a concrete model of quasi-single field inflation \cite{Chen:2009we}. We leave the alternative-to-inflation models to future study.

In this model, besides the universal gravitational coupling, the inflaton is coupled to a massive field with mass of order $H$ through a direct derivative coupling, and the massive field also has non-linear self-couplings. Following the notations in \cite{Chen:2009we}, we denote the sizes of these couplings as $\theta_0/H$ and $V'''/H$ (for the cubic coupling), respectively. Both couplings are required to be much less than one due to the limitation of the perturbative calculation.
The natural embedding of this model in the supergravity theory and its generalization are shown in \cite{Baumann:2011nk}.
The amplitude of the three-point function due to a variety of couplings have been estimated in the Appendix A of \cite{Chen:2009we} and in \cite{Baumann:2011nk}.
We comment that, in the diagram that contains the massive field self-coupling, the clock signal is still generated in the same qualitative way between the two short massless modes and one long massive mode, as we described in Sec.~\ref{Sec:Quantum_massive} and \ref{Sec:QSC}. The intermediate propagators of short massive modes do not spoil the resonance, because the time integration contains a region in which the phases of the massive mode and its complex conjugate cancel each other in the propagator. We can also directly use the amplitude estimates in \cite{Chen:2009we,Baumann:2011nk} to get some ideas on the range of the size of the clock signals, since it is a smooth analytic continuation across the $m=3H/2$ point.

\begin{itemize}

\item
If the direct coupling between the inflaton and the clock field is turned off and there is only the universal gravitational coupling, the clock-field-mediated non-Gaussian curvature scalar correlations appear as loop diagrams, since the $\delta\sigma$'s always appear in pairs in the vertices of the gravitational coupling. The amplitude of these correlation functions are suppressed by factors of $\epsilon$ and $P_\zeta\sim 10^{-9}$. For example, using the cubic and quartic couplings in \cite{Chen:2012ye}, we can estimate that for the bispectrum $f_{\rm NL}$ is proportional to $\epsilon P_\zeta$ with some possible enhancement factors from the UV cutoff scale, which may compensate the $P_\zeta$ suppression in non-Gaussianity.

\item
Considering the direct coupling between the inflaton and the clock field through the derivative coupling turns on a bilinear coupling between them. Depending on whether the three-point vertex is of the type $\zeta^2 \delta\sigma$ or $\zeta \delta\sigma^2$ and the derivative structures in these vertices, the amplitude of the bispectrum can be $f_{\rm NL} \propto \CO(\epsilon (\dot\theta/H)^2)$, $\CO(\epsilon)$, or $\CO((\dot\theta/H)^2)$. Here $\epsilon$ is $\CO(0.01)$ or much less, and $(\dot\theta/H)^2\ll 1$ for perturbative calculation.

\item
Considering self-interactions of the clock field turns on a larger source of non-linearity, and the resulting bispectrum has an amplitude
\bea
f_{\rm NL} \propto P_\zeta^{-1/2} (V'''/H) (\dot\theta_0/H)^3 \sim 10^5 (V'''/H) (\dot\theta_0/H)^3 ~,
\label{fNL_self}
\eea
again $V'''/H$ and $\dot\theta_0/H$ are much less than one.
This possibility gives the largest non-Gaussianity among the examples mentioned here.
Note that the amplitude of $f_{\rm NL}$ here is evaluated at the equilateral configuration. For the squeezed configurations, we also have to take into account the envelop behavior in \eqref{Clock_signals} and an extra factor of $k_3/k_1$ due to the reduction of configuration space when detectability is concerned, both are suppression factors for the inflation scenario.
Also note that the amplitude of \eqref{fNL_self} is mostly independent of whether the inflation model is of the large or small field type.

\item
As the mass of the clock field becomes much larger than the Hubble parameter $H$, all the amplitudes of the clock signals above are expected to be suppressed by a Boltzmann factor. This factor should be roughly of the form $\sim \exp(-\pi m/H)$ as in \cite{Arkani-Hamed:2015bza}, and the details are under investigation. Therefore, for the observational purpose, the most important standard clock fields should be those with mass larger than $3H/2$ but still of order $H$.
Interestingly, scalar fields with mass of order $H$ are generically expected in inflation models in supergravity.

\end{itemize}

As we can see, the amplitude of the clock signals spans a wide range and is very model-dependent.
While in many cases the signals are small,
some of them, including some natural ones in the supergravity model-building, may fall into the observable ranges of the CMB, LSS, and 21 cm experiments. It would be interesting to study the observability of the signals in future experiments. (See Ref.~\cite{Sefusatti:2012ye,Norena:2012yi} for a forecast on the non-oscillatory signals in quasi-single field inflation models.) We also note that, due to the special shape-dependent oscillation patterns with variables such as frequency and phase, these signals are most likely orthogonal to the existing popular templates used in non-Gaussianity data analysis. So dedicated templates and data analyses are necessary to constrain or discover them, for example, by applying the method of modal decomposition \cite{Fergusson:2006pr}.

\section{Conclusions and discussions}
\setcounter{equation}{0}
\label{Sec:Discussions}

Cosmological observables that can be used to model-independently distinguish the inflation scenario from possible alternative scenarios are very rare and an important subject of theoretical and experimental study.
In this paper, motivated by the development in the classical primordial standard clock models \cite{Chen:2011zf,Chen:2014cwa} and quasi-single field inflation models \cite{Chen:2009we,Arkani-Hamed:2015bza}, we pointed out and studied a novel type of such signals generically existing in the primordial universe models, independent of the well-known tensor mode approach. These signals are generated by quantum fluctuating massive fields, and are the quantum mechanical realization of the idea of the primordial standard clocks. The ticks of the massive field oscillations get imprinted as oscillatory features in the shape-dependence of various primordial non-Gaussianities, which directly encode the time-dependence of the scale factor of the primordial universe $a(t)$. We have demonstrated how this is achieved through qualitative arguments as well as detailed examples in both inflation and alternative-to-inflation scenarios. In the inflation example, we are able to derive several exact results. In the alternative-to-inflation examples, we are able to use the universal features of the standard clocks to derive some general results.

Here to highlight some key properties, we summarize the main differences between the classical standard clocks, studied in previous works, and the quantum standard clocks of this work.

\begin{itemize}
\item
Classical standard clocks exist only if some massive fields get excited {\em classically} by certain background sharp features. In contrast, the quantum standard clocks generically exist in realistic primordial universe models because massive fields always fluctuate {\em quantum mechanically} in time-dependent background. Depending on the structures of couplings in detailed models, the signals may show up in different types of three-point and/or higher order correlation functions, with the universal gravitational coupling case being the minimum case.
However, the amplitude of the signals is highly model-dependent. This situation is much like that for the tensor mode.

\item
Due to the generation mechanism, the simplest classical standard clock models predict two different kinds of signals that are smoothly connected to each other, namely the {\em sharp feature signal and clock signal}; while the simplest quantum standard clock models only have the {\em clock signal}.

\item
Classical standard clocks break the scale invariance. The clock signals manifest as {\em scale-dependent} oscillatory features in various correlation functions including the power spectrum and non-Gaussianities. In contrast, the quantum standard clock itself is not a source of scale-invariance breaking,\footnote{
This does not include the possible scale-dependence which may be introduced by the primordial universe model itself when clock fields are ignored, especially in the alternative-to-inflation scenarios.
Relatedly we note that, for alternative-to-inflation scenarios, if we rescale all momenta in \eqref{Clock_singals_1} and \eqref{Clock_signals}, the phases may not be invariant at the leading order due to the possible scale-dependence in $H_{k_3}$ and $m_{{\rm h},k_3}$. Such a dependence does not encode the clock signal.
}
because it appears persistently due to its quantum nature. So, the clock signals are absent in the power spectrum, and are only present in non-Gaussianities and manifest as {\em shape-dependent} oscillatory features.

\end{itemize}

There are many important directions that need to be studied for the quantum primordial standard clocks. For example,

\begin{itemize}

\item
We have assumed that the mass of the clock field remains constant. Time-dependent mass terms are naturally induced in the background where there is no shift symmetry in time.
If the mass of the field is much larger than the horizon scale, this time-dependent contribution to the mass is negligible. This is usually the case for the classical standard clocks, in which the amplitude of the clock signal increases as the mass of clock increases \cite{Chen:2011zf,Chen:2014cwa}. However, for the quantum standard clock models, the signal is exponentially suppressed by a Boltzmann factor for heavy clock fields, so the observable range is where the mass of the clock field is not too much larger than the horizon scale. The time-dependence in the mass term mentioned above can become important in this case, and it is interesting to study how it affects the predictions.

\item
We have focused on the shape of non-Gaussianities, but not their amplitudes. The main point is to show the general existence and physical significance of the quantum standard clock. Nonetheless, the amplitude is critical for the observability of such signals, and as discussed it is highly model-dependent. A promising example is the example of quasi-single field inflation model studied in \cite{Chen:2009we}. It will be interesting to extend the results in \cite{Chen:2009we} to $m>3H/2$ and compute explicitly the clock signal in this model. This could provide a concrete inflation model example in which the quantum standard clock signal can be observably large.

\item
We only considered one type of the three-point function with specific interaction vertex  depicted in Fig.~\ref{Fig:3pt_vertex}. The clock signals are also encoded in shapes of other types of three-point functions (with different vertices) and higher-order correlation functions, in terms of similar squeezed configurations or other special momentum configurations. This is another important direction. Related studies in quasi-single field inflation models can be found in \cite{Assassi:2012zq,Arkani-Hamed:2015bza}.

\item
In the analysis of the alternative-to-inflation scenarios, we have focused on generic properties of the quantum standard clocks that are insensitive to model-building details. It would be interesting to study more explicit and concrete models.

\item
In this work, we have emphasized the key theoretical predictions and physical significance of the quantum standard clocks. An important future step is to survey different models, come up with simple templates for such signals and implement them in the actual data analyses.

\end{itemize}

\medskip
\section*{Acknowledgments}

We thank Nima Arkani-Hamed, Daniel Eisenstein, Avi Loeb, and Juan Maldacena for helpful discussions.
XC and MHN are supported in part by the NSF grant PHY-1417421.
YW is supported by Grant HKUST4/CRF/13G issued by the Research Grants Council (RGC) of Hong Kong.
Part of this work was performed at the Aspen Center for Physics during the workshop ``Primordial Universe", which is supported by National Science Foundation grant PHY-1066293.

\appendix

\section{Clock signal and (non-)resonant integral}
\label{App:resonance}
\setcounter{equation}{0}

As we explained in the main text, the ticks of the massive clock field oscillation get imprinted in the density perturbations and record the time-evolution of the background scale factor. This process is mostly done through the resonance mechanism \cite{Chen:2008wn,Flauger:2009ab,Flauger:2010ja,Chen:2010bka}. In order for the resonance mechanism to take place, if the massive field
contribution to the integrand behaves as $\sim e^{\mp imt}$, then the curvature scalar field contribution has to behave as $\sim e^{\pm iK\tau}$.
However, as we have seen in Sec.~\ref{Sec:non-T-ordered}, in the inflation case, the clock signal appears even if the two modes in certain integral do not have the required arrangement above and hence do not resonate with each other. In such a case, the resulting clock signal is suppressed by an exponentially small factor; but nonetheless it may still be important because such a non-resonant integral may carry an exponentially larger weight than the resonant one, due to some natural initial conditions. In Sec.~\ref{App:resonance_inflation}, we illustrate this point in terms of simple integrals.

We also study the same issue for the alternative-to-inflation models with $0<p<1$. In Sec.~\ref{App:resonance_alter}, through several representative examples we observe that, unlike the inflation case, resonance is required to generate the clock signal. However we do not have a general proof for arbitrary values of $p$ within $0<p<1$, and so cannot exclude the possible existence of special cases. This conclusion will be used in Sec.~\ref{Sec:Alternatives}.

\subsection{Exponential inflation example}
\label{App:resonance_inflation}

The integral that we refer to in the inflation case is the term in the 2nd bracket of (\ref{Term1}):
\bea
c_3 \frac{\sqrt{\pi}}{4} H \frac{(k_1k_2)^{1/2}}{k_3^{5/2}} e^{-\frac{\pi}{2}\mu}
\int_{-\infty}^0 dx (-x)^{3/2} H^{(1)*}_{i\mu}(-x) e^{i\frac{k_1+k_2}{k_3}x} ~.
\label{term1_Appendix}
\eea
This definite integral is done exactly in Sec.~\ref{Sec:non-T-ordered}. For example, in the very squeezed limit, $k_3/k_1\ll 1/\sqrt{\mu}$, and the large mass limit, $\mu \gg 1$, we have
\bea
c_3 \frac{\sqrt{\pi}}{4} H \mu^{3/2} e^{-\pi\mu} (2k_2)^{-3/2}
\left[ i\left( \frac{4k_1}{k_3} \right)^{-i\mu}
+ \left( \frac{4k_1}{k_3} \right)^{i\mu} \right] ~.
\label{Term1_inflation}
\eea
We have chosen the massive field $v_k$ to be originated from the BD vacuum. In the classical regime, this choice (after complex conjugation) leads to an oscillatory behavior that is a linear combination of the negative and positive frequency, $(-x)^{-i\mu}$ and $(-x)^{i\mu}$; but with very different weights -- the former is a factor of $e^{\pi\mu}$ larger than the latter. The positive frequency component $\sim (-x)^{i\mu}$ resonates with the massless mode $\sim e^{i\frac{k_1+k_2}{k_3}x}$ in the integrand, giving a clock signal which is the first term in (\ref{Term1_inflation}). Interestingly, although the negative frequency component does not resonate, the convoluted integration with the massless mode contributes a term that also contains a similar clock signal (the 2nd term in (\ref{Term1_inflation})); there is much cancellation during this integration due to the absence of resonance, but this is compensated by the large weight of this component. In the end, the two clock signals in (\ref{Term1_inflation}) are equally important, causing some phase shift to each other.
We also note that, in some integrals such as $\cal{I}_{\rm Hankel}$ in \eqref{I2_def}, the massive field mode function differs from that in \eqref{term1_Appendix} by a complex conjugate. Then the component that resonates with the curvature scalar is different, and the non-resonant clock signal becomes negligible.

To best illustrate the point, we look at the following toy integral,
\ba
\calI_\pm (p) =\int^0_{-\infty} d\tau e^{\pm imt} e^{-iK\tau} ~,
\label{toy_integral}
\ea
where ${\cal I}_+$ corresponds to the resonant integral, and $\calI_-$ the non-resonant integral.
This integral captures the essential behaviors of both oscillatory terms in \eqref{term1_Appendix}, and for the purpose of this appendix, we ignore the other slowly varying functions in the integrand.

In the case of exponential inflation that we consider in this paper, $p\to \infty$, so the integral is in the following form
\ba
\calI_\pm(\infty) = \int_{-\infty}^0 (-\tau)^{\pm i \mu} e^{-i K \tau} ~.
\ea
After using the $i\epsilon$ prescription to make the integrals UV convergent, one can obtain the following exact results
\ba
\calI_+ (\infty) = (-i K)^{-1+i \mu }\, \Gamma (1-i \mu )
\\
\calI_- (\infty) =(-i K)^{-1-i \mu } \, \Gamma (i \mu +1).
\ea
It is clear that both resonant and non-resonant integrals give the clock signal.\footnote{For cases where $p > 1$ but still at the order of one, the situation is a little tricky. For example, we have checked that for $p = 2$ the non-resonant integral does not produce the clock signal while for $p = 4$ it does but with a different frequency from the resonant case. The more non-trivial case is $p=3$ where the resonant and non-resonant integrals give exactly the same results and there are two oscillatory factors with different frequencies (though the scale dependence of the frequencies is the same, consistent with the clock signal).}
However, the non-resonant integral is suppressed in comparison with the resonant one. This is indeed expected as the resonance enhances the amplitude, while the non-resonant integral is suppressed due to the highly oscillatory factors. To see this fact more clearly, it is useful to expand the exact results in large mass limit, i.e. $\mu \gg 1$. One has
\ba
\calI_+ (\infty) &\xrightarrow{\mu \gg 1}& (1+i) \sqrt{\pi } \, e^{i \mu } \, \mu ^{\frac{1}{2}-i \mu }\, K^{-1+i \mu }
\\
\calI_- (\infty) &\xrightarrow{\mu \gg 1}& (-1+i) \sqrt{\pi } \,  e^{-(\pi +i) \mu } \, \mu ^{\frac{1}{2}+i \mu }\, K^{-1-i \mu }.
\ea
The extra factor of $e^{-\pi \mu}$ in $\calI_-$ is responsible for the suppression of the non-resonance integral.

\subsection{Alternative to inflation ($0<p<1$)}
\label{App:resonance_alter}

In this section, to study the same issue in the contracting scenarios ($0<p<1$), we use the same toy integral \eqref{toy_integral}.
Following the same convention in Sec.~\ref{Sec:Alternatives}, for $0<p<1$ this integral becomes
\ba
\calI_\pm (p) = \int^0_{-\infty} d\tau \, e^{\mp i(1-p) \nu (-\tau)^{\frac{1}{1-p}}} e^{-iK\tau} ~.
\label{I_alt}
\ea
We are unable to analytically integrate \eqref{I_alt} with general $p$, so, in the following we compute this integral with representative $p$ values for two types of contraction scenarios, respectively. The conclusion applies to these examples at least, although we do not have a general proof.

\subsubsection{matter contraction ($p=2/3$)}
In this case we have
\ba
\calI_+(2/3) &=&
-\frac{i K^5}{40 \nu ^2} \, _1F_4\left(1;\frac{7}{6},\frac{4}{3},\frac{5}{3},\frac{11}{6};\frac{K^6}{1296 \nu ^2}\right)
+\frac{i K^2}{2 \nu } \, _1F_4\left(1;\frac{2}{3},\frac{5}{6},\frac{7}{6},\frac{4}{3};\frac{K^6}{1296 \nu ^2}\right)
\nonumber \\
&&+\frac{\pi}{3 \nu^{1/3}}  \left(3 \text{Ai} (-\frac{K}{\nu^{1/3}} )-i \, \text{Bi} (-\frac{K}{\nu^{1/3}})\right)
\label{I+2/3}
\ea
and
\ba
\calI_-(2/3) &=&
-\frac{i K^5}{40 \nu^2} \, _1F_4 \left(1;\frac{7}{6},\frac{4}{3},\frac{5}{3},\frac{11}{6};\frac{K^6}{1296 \nu ^2} \right)
-\frac{i K^2}{2\nu} \, _1F_4\left(1;\frac{2}{3},\frac{5}{6},\frac{7}{6},\frac{4}{3};\frac{K^6}{1296 \nu ^2}\right)
\nonumber \\
&&+\frac{\pi}{3 \nu^{1/3} }  \left(3 \text{Ai} (\frac{K}{\nu^{1/3}} )+i \, \text{Bi} (\frac{K}{\nu^{1/3}} )\right) ~,
\ea
where $_1F_4$ is the Generalized Hypergeometric function and $\text{Ai}$ and \text{Bi} are Airy functions.
Note that one can reproduce $\calI_-$ from $\calI_+$ by the analytic continuation $\nu \to \nu e^{-i\pi}$.
In order to see whether or not the results show the clock signal it is useful to expand the results for large arguments, where we expect characteristic oscillatory behavior in the clock signal:
\ba
\calI_+(2/3) & \xrightarrow{K \gg 1}&
\propto
\left(\sqrt{4+3 i} \, \sin  (\frac{2 K^{3/2}}{3 \sqrt{\nu }} )+\sqrt{4-3 i}\, \cos  (\frac{2 K^{3/2}}{3 \sqrt{\nu }} ) \right)
\label{I+2/3_largeK}
\\
\calI_-(2/3)  &\xrightarrow{K \gg 1}&
\frac{\sqrt{\pi }}{2(K\nu)^{1/4}} e^{-\frac{2 K^{3/2}}{3 \sqrt{\nu }}}
+ \frac{i}{K}
~,
\ea
where we kept the leading order terms in the real and imaginary parts. In \eqref{I+2/3_largeK} we could not determine the precise prefactor directly from \eqref{I+2/3}. The above expansions show clearly that the resonant integral gives the clock signal expected from the saddle point approximation, while the non-resonant integral is just a decaying function in $K$. Note that the above approximate results are {\it not} related by the analytic continuation $\nu \to \nu e^{-i\pi}$. That is, the expansion for large $K$ and the analytic continuation do not commute.

\subsubsection{A slowly contracting scenario (p=1/5)}
We next study the same integral in a slowly contracting scenario with $p=1/5$. Again the integrals can be computed exactly and one has
\ba
\calI_+ (1/5) &=& \frac{5^{4/5}}{(-2)^{8/5} \nu ^{4/5}}\ \Gamma (\frac{9}{5}) \ _3F_3\left(\frac{9}{20},\frac{7}{10},\frac{19}{20};\frac{2}{5},\frac{3}{5},\frac{4}{5};\frac{i K^5}{5 \nu ^4}\right)
\\ \nonumber
&-&\frac{ (-1)^{9/10} \ 5^{11/5}}{96\ 2^{2/5} \nu ^{16/5}}  K^3 \ \Gamma (\frac{16}{5} ) \ _3F_3\left(\frac{21}{20},\frac{13}{10},\frac{31}{20};\frac{6}{5},\frac{7}{5},\frac{8}{5};\frac{i K^5}{5 \nu ^4}\right)
\\ \nonumber
&-&\frac{7 (-1)^{4/5}}{4\ 2^{4/5} 5^{3/5} \nu ^{12/5}} K^2\ \Gamma (\frac{2}{5}) \ _3F_3\left(\frac{17}{20},\frac{11}{10},\frac{27}{20};\frac{4}{5},\frac{6}{5},\frac{7}{5};\frac{i K^5}{5 \nu ^4}\right)
\\\nonumber
&-&\frac{3 (-1)^{7/10} }{2^{6/5} 5^{2/5} \nu ^{8/5}} K \, \Gamma (\frac{3}{5}) \, _3F_3\left(\frac{13}{20},\frac{9}{10},\frac{23}{20};\frac{3}{5},\frac{4}{5},\frac{6}{5};\frac{i K^5}{5 \nu ^4}\right)
\\ \nonumber
&+& \frac{125 }{256 \nu ^4}K^4 \, _4F_4\left(1,\frac{5}{4},\frac{3}{2},\frac{7}{4};\frac{6}{5},\frac{7}{5},\frac{8}{5},\frac{9}{5};\frac{i K^5}{5 \nu ^4}\right) ~,
\ea
\ba
\calI_- (1/5) &=& \frac{(-2)^{2/5}}{5^{1/5} \nu ^{4/5}} \Gamma (\frac{4}{5}) \ _3F_3\left(\frac{9}{20},\frac{7}{10},\frac{19}{20};\frac{2}{5},\frac{3}{5},\frac{4}{5};\frac{i K^5}{5 \nu ^4}\right)
\\ \nonumber
&+&\frac{125}{256 \nu ^4} K^4 \  _4F_4\left(1,\frac{5}{4},\frac{3}{2},\frac{7}{4};\frac{6}{5},\frac{7}{5},\frac{8}{5},\frac{9}{5};\frac{i K^5}{5 \nu ^4}\right)
\\\nonumber
&-&\frac{3 (-1)^{3/10}}{2^{6/5}\ 5^{2/5} \nu ^{8/5}} K \ \Gamma (\frac{3}{5}) \ _3F_3\left(\frac{13}{20},\frac{9}{10},\frac{23}{20};\frac{3}{5},\frac{4}{5},\frac{6}{5};\frac{i K^5}{5 \nu ^4}\right)
\\ \nonumber
&-&\frac{25 (-1)^{1/10} 5^{1/5}}{96\ 2^{2/5} \nu ^{16/5}} K^3 \ \Gamma (\frac{16}{5} ) \ _3F_3\left(\frac{21}{20},\frac{13}{10},\frac{31}{20};\frac{6}{5},\frac{7}{5},\frac{8}{5};\frac{i K^5}{5 \nu ^4}\right)
\\ \nonumber
&+&
\frac{7 (-1)^{1/5}}{4\ 2^{4/5} 5^{3/5} \nu ^{12/5}} K^2 \ \Gamma(\frac{2}{5}) \, _3F_3\left(\frac{17}{20},\frac{11}{10},\frac{27}{20};\frac{4}{5},\frac{6}{5},\frac{7}{5};\frac{i K^5}{5 \nu ^4}\right) ~.
\ea
Expanding the above exact results in large $K$ limit one has
\ba
\calI_+(1/5) & \xrightarrow{K \gg 1}&
\frac{2(1- i) \sqrt{\pi }}{\nu ^2} K^{3/2} \ e^{\frac{i K^5}{5 \nu ^4}} ~,
\\
\calI_-(1/5)  &\xrightarrow{K \gg 1}&
\frac{i}{K}+
\frac{4 (-1)^{5/8} \, 5^{3/4} \, \pi ^2 \nu  \, \Gamma (\frac{1}{4})}{ \Gamma (-\frac{1}{20} ) \Gamma  (\frac{3}{20} ) \Gamma  (\frac{7}{20} ) \Gamma  (\frac{11}{20})} \, \dfrac{1}{K^{9/4}} ~,
\ea
where we have kept the leading order terms in the imaginary and real parts. Again, only the resonance integral is oscillating and the frequency of oscillation is consistent with the clock signal obtained by using the saddle point approximation.

We have checked that the same conclusion applies to the $p=1/10$ case.

\end{spacing}

\end{document}